# Bell-type Models and Statistical Description of Quantum Systems

Yu.I. Bogdanov

*Institute of Physics and Technology, Russian Academy of Sciences*[1]

We consider dynamics of hidden variables for measurements in a generalized bell-type model for a single spin using natural assumptions. The evolution of the system, which can be expressed as dynamic chaos is studied. The equilibrium state that the system evolves to asymptotically is consistent with the predictions of quantum theory. The thesis of incompleteness of quantum mechanics in dynamic interpretation and completeness in statistical interpretation are developed. Conceptual problems of quantum mechanics such as violations of Bell inequalities, negative probabilities, complementarity principle, Einstein's locality and others are discussed.

## Introduction

Statistical nature of quantum mechanics is one of the most important of its properties. The impossibility of a deterministic prediction of quantum events raised concerns in completeness of quantum theory and became the basis for hypothesis of existence of hidden variables that would allow statistical descriptions to provide deterministic predictions. According to a common framework, it is impossible to introduce such hidden variables into quantum theory. At the same time, it appears that researchers often use models with hidden variables without even understanding the fact.

For instance, let us consider the problem of modeling quantum systems on a classical deterministic computer. In this problem, every representative of a quantum ensemble is inevitably characterized not by a quantum state $\psi$, but by a pair ($\psi$, $\lambda$), where $\lambda$ - is an individual hidden variable for given representative, which is usually chosen at random. It is impossible to imitate a statistical quantum experiment without introducing such variable. Actually, if the state of two electrons is the same $\psi$, then why in the Stern-Gerlach device should one of the electrons move upwards and the other downwards?

The simulation statistical models considered are a transition of a well known method of Monte-Carlo into quantum events. It seems that such models should not be considered as the actual dynamic theories that would allow to move from quantum mechanics to some deeper and more fundamental description of the Nature. However, historically it was the case. It will be demonstrated that Bell in his 1966 work tried to reject von Neumann's proof of impossibility of introduction of hidden variables by proposing a counter-example of such a model that reduced to simulation modeling of statistical distributions. In fact, Bell considers simulation (by means of a hidden generator of random numbers) of data of measuring spin ½ that leads to Bernoulli distribution. Note, that Bell uses the language of "quantum philosophy" and not computer modeling and thus he does not make a conclusion that his model completely leads to the well known Monte-Carlo method. At the same time, Bell's arguments do not embed any logical inconsistencies - a similar model for spin ½ was proposed in 1967 by Cochen and Specker. The aforementioned works were widely accepted as having apparent descriptions of hidden variables for quantum states of a spin-½ particle. Note that there are significant differences in the ways problems are formulated for Hilbert spaces of dimension two (spin ½) and those of higher dimension

Thus, the Bell model can be considered either as a philosophical deterministic issue or an applied statistical simulation of a quantum state. The choice depends on the context of the problem. At the same time, the construction of the actual Bell's system can be performed in many different ways – as an abstract quantum bit (qubit), ½-spin particle, photon's polarization degree of freedom, two-level atom, etc.

[1] E-mail: bogdanov@ftian.ru

Note also, that Bell does not discuss the dynamics of hidden variables, as he restricts his model to a single measurement. However, this question is of most importance if considered in the framework of quantum mechanics. For example, if we consider that hidden variables do not change between measurements, then we get to the classical urn scheme, which does not take into account some extremely important quantum phenomena, such as the existence of non-commuting variables.

Therefore, in accordance with the aforementioned facts, the problem of statistical simulation should be formulated as follows. A large number $n$ of representatives of a quantum statistical ensemble are given as inputs to the simulation algorithm. The state of each representative is given by a pair $(\psi, \lambda_i), i = 1,...,n$, where $\psi$ has the same value for all representatives and $\lambda_i$ is a individual variable for each $i$. Then the natural question is - how should one set the input values of $\lambda_i$ and what their behavior in the numerical algorithm should be like for a meaningful modeling of quantum systems?

The answer is the following (detailed explanation is to be given this paper). The input values $\lambda_i$ should be defined by a uniform random number generator. At the same time the dynamics of the hidden variables is described according to the rule – it may be considered that $\lambda_i$'s do not change for unitary transformations, while during measurements they are subject to some uniform extension transformation. Then the results of a computer simulation will be identical to the results of the actual physical experiments. If that is the case, can we consider the real physical experiments as some simulation with hidden variables performed by the Nature? Thus, as we can see, there is no gap between the philosophical and pragmatic approaches.

## 1. Hidden variables and deterministic imitation of quantum states

### 1.1 Bell's model of hidden variables for a ½-spin particle – determinism instead of probability.

For the simple case of a quantum state in a 2D-Hilbert space, an explicit model containing hypothetical hidden variables was proposed by Bell. The model was designed to substitute a statistical description to a deterministic prediction regarding the results of a quantum measurement [1].

Below we shall describe the Bell's model together with obvious intermediate calculations. Bell considers the quantum state of a ½-spin particle. A further assumption of independence between the spin and translational degrees of freedom is made. Then the wave function of the particle is the product of a spin function and a coordinate wave function, which allows us to consider the spin state independently from the coordinate. In modern terminology, we may say that the spin and the coordinate of a particle are not entangled with each other. The considered system is set in a Hilbert space of dimension two. The quantum state of such particle is a spinor $\psi = \begin{pmatrix} \psi_1 \\ \psi_2 \end{pmatrix}$.

Any observable in the system is a Hermitian matrix 2x2 that may be expressed as: $\hat{Y} = \alpha + \vec{\beta}\vec{\sigma}$, where $\alpha$ - is a real number, $\vec{\beta} = (\beta_x, \beta_y, \beta_z)$ - is a real vector and, $\vec{\sigma}$ - are Pauli matrices.

Measurements of the observable $\hat{Y}$ will lead to one of the eigenvalues: $\alpha \pm |\vec{\beta}|$.

We shall suppose, without loss of generality, that the axis $Z$ is defined parallel to the spin polarization. In this system of coordinates, a spin state is: $\psi = \begin{pmatrix} 1 \\ 0 \end{pmatrix}$. It is simple to calculate the mean value of the observable:

$$M(\hat{Y}) = \langle \psi | \alpha + \vec{\beta}\vec{\sigma} | \psi \rangle = \alpha + \beta_z ,$$

where $\beta_z$ - is a projection of vector $\vec{\beta}$ on axis $z$ .

Then Bell's argument proceeds as follows. If we introduce some hypothetical hidden variable $\lambda$ , then we may transform statistical predictions of quantum mechanics into completely determined ones. Following Bell's argument, in order to do so, we need to define the state of the system using not only spinor $\psi$ but also variable $\lambda$, which is a random value uniformly distributed at [–0.5 0.5] (i.e. $-0.5 \le \lambda \le 0.5$).

According to Bell, the value of observable $\hat{Y}$ can de defined deterministically as:

$$Y(\lambda) = \alpha + |\vec{\beta}|\, sign\left(\lambda|\vec{\beta}| + \frac{1}{2}|\beta_z|\right) signX ,$$

where $X = \begin{cases} \beta_z, & \beta_z \ne 0 \\ \beta_x, & \beta_z = 0, \beta_x \ne 0 \\ \beta_y, & \beta_z = 0, \beta_x = 0 \end{cases}$

The sign function is defined in the usual manner:

$$signX = \begin{cases} +1, & X \ge 0 \\ -1, & X < 0 \end{cases}$$

The observable $Y(\lambda)$ depends on the hidden variable and thus it takes one of the two values $\alpha \pm |\vec{\beta}|$. It is not difficult to show that the uniform distribution of the variable $-0.5 \le \lambda \le 0.5$ leads to the correct mean value of operator $\alpha + \vec{\beta}\vec{\sigma}$ , equal to $\alpha + \beta_z$, which is consistent with quantum mechanics:

$$M(Y) = \int_{-0.5}^{0.5} Y(\lambda) d\lambda = \int_{-0.5}^{0.5} \alpha + |\vec{\beta}|\, sign\left(\lambda|\vec{\beta}| + \frac{1}{2}|\beta_z|\right) signX d\lambda = \alpha + \beta_z$$

As the result, the mean value of operator $\hat{Y}$ in quantum mechanics is equal to the mean value of the random observable $Y$. Thus, introduction of a hidden variable $\lambda$, completely eliminates quantum uncertainty and results in a deterministic prediction.

Note, however, that Bell considers only mean values and does not take into account the probabilities explicitly. We shall complete Bell's calculations by explicitly calculating the corresponding probabilities to understand the nature of the hidden variable $\lambda$ in more detail.

**1.2. Geometric interpretation of probability**

Simple calculations yield the following well-known result – the eigenvalue $\alpha + |\vec{\beta}|$ (spin upwards) will be observed with probability $\cos^2(\theta/2)$, where $\theta$- is the angle between vector $\vec{\beta}$ and axis $Z$ ; by analogy the probability of $\alpha - |\vec{\beta}|$ (spin downwards) is $\sin^2(\theta/2)$.

The mean value of observable $\alpha + \vec{\beta}\vec{\sigma}$ is:

$$\left(\alpha + \left|\vec{\beta}\right|\right)\cos^2(\theta/2) + \left(\alpha - \left|\vec{\beta}\right|\right)\sin^2(\theta/2) = \alpha + \beta_z$$

It is easy to see that the procedures for measurements of spin on axis $\vec{\beta}$ and $-\vec{\beta}$ physically coincide (positive spin on axis $\vec{\beta}$ is interpreted as the negative on axis $-\vec{\beta}$ and vice versa). Thus, we may limit ourselves to considering only the upper hemisphere ($0 \le \theta \le \pi/2$).

The essence of the Bell's model can be explained as follows. We may associate complete probability with the segment of unitary length ($-0.5 \le \lambda \le 0.5$). The position of the point on the segment is predicted by quantum mechanics, and the interval is separated in two parts with weights $\sin^2(\theta/2)$ and $\cos^2(\theta/2)$ (complete probability is derived as the sum of the considered weights, just as the interval does). According to the elementary concept of geometric probabilities [2], the Bell's hidden variable $\lambda$ merely completes statistical prediction of quantum mechanics to deterministic one. If the value of $\lambda$ lies on the right side of length $\cos^2(\theta/2)$, then one will detect upward spin (eigenvalue $\alpha + \left|\vec{\beta}\right|$); otherwise the spin will be detected downwards (eigenvalue $\alpha - \left|\vec{\beta}\right|$).

### 1.3. Hidden parameter and probability distribution function

It is clear that the Bell's hidden variable merely reflects the impossibility of statistical prediction of the behavior of a given representative of an ensemble. In this sense we may state incompleteness of any statistical description. Such incompleteness always implicitly exists in probability theory and mathematical statistics.

In mathematical statistics, Bell's hidden variable (or parameter $F = \lambda + 0.5$, to be more precise) is the value of so-called cumulative distribution function. It is a well known fact that the value this function is uniformly distributed on [0 1].

This measurement can be described as a binomial distribution. The corresponding cumulative distribution function is shown on the Fig.1:

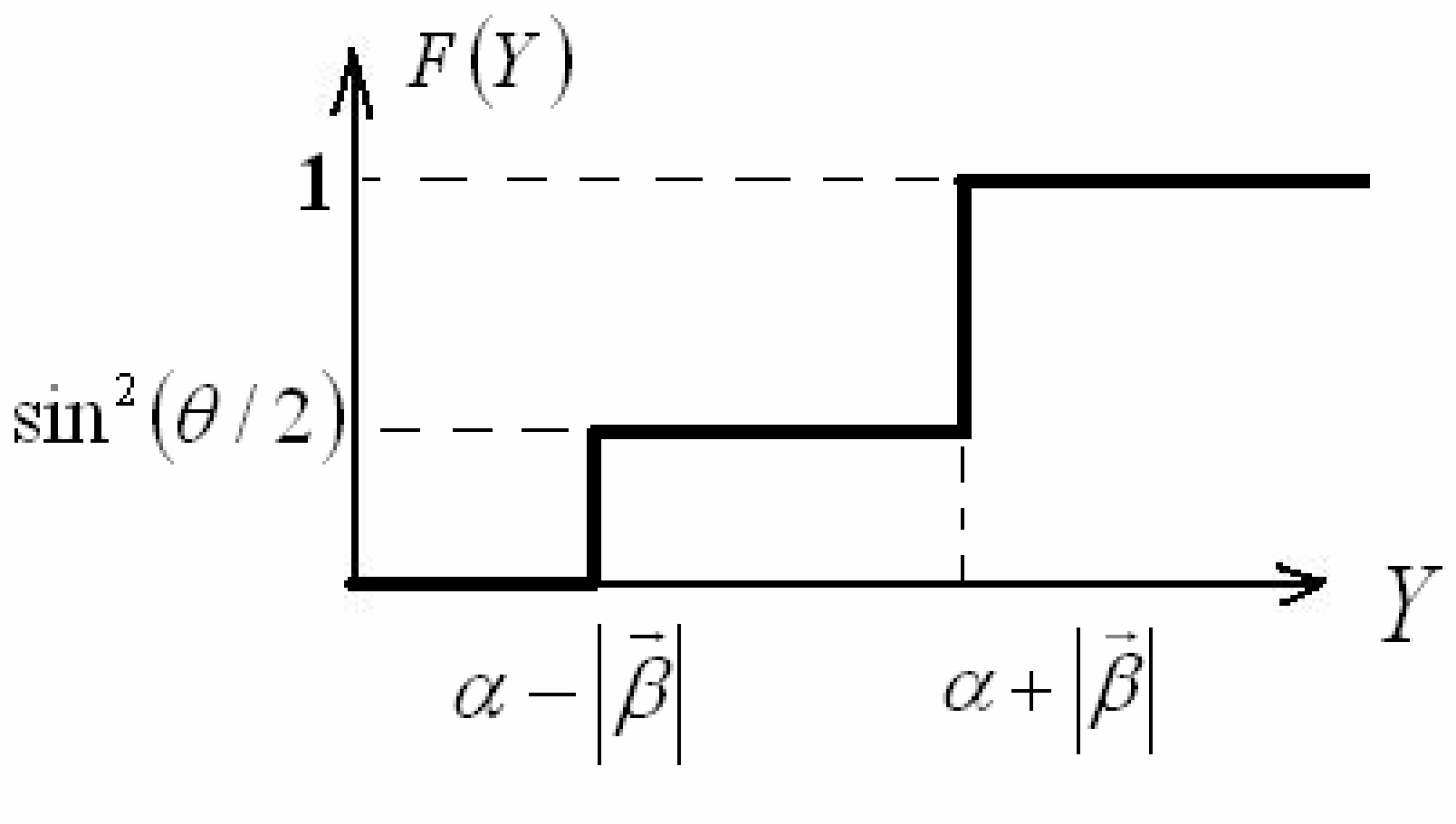

Figure 1. Cumulative distribution function for a random value corresponding to observable $\hat{Y}$

### 1.4. Bell model and Monte-Carlo method.

The previous discussion demonstrates that the essence of the Bell's approach is actually the well-known Monte-Carlo method, which allows for statistical sample generation. The method is widely used for statistical modeling of both classical and quantum systems (in fact, real statistical models are substituted by Bell-type models with hidden variables). Thus, the Bell's hidden variable is simply a generator of a uniform distribution (random number generator). It is well known that if the value generated by a random number generator is a priori unpredictable (hidden) and is uniformly distributed, then the results of Monte-Carlo modeling are indistinguishable from results of the real statistical experiment. For example, a real statistical coin experiment is equivalent to modeling using random data generator.

Real experiments with ½-spin particles are quantum realization of a coin experiment, where the coin is not necessarily fair but rather has the probability of $\cos^2(\theta/2)$ of one side flip and the angle $\theta$ is tuned in the experiment. All the aforementioned considerations were clearly well-known to von Neumann, who was one of the founders of Monte-Carlo method. This is important to note because the Bell's model was primarily aimed at criticizing the von Neumann's approach.

Thus, one comes to the conclusion that the substitution of statistical model of quantum mechanics by a deterministic model with hidden variables is simply a way to model statistical distributions using random data generator (this approach is well-known and is equivalent to the Monte-Carlo approach)

### 1.5. Gödel -type incompleteness

Basing upon the elementary analysis above, one may come to the conclusion that the Bell model is trivial and thus does not deserve much attention. This, however, is not true. The model became a significant milestone marking the differences between two prevailing attitudes on the possibility of deterministic description of purely statistical phenomena. Early on, until the development of quantum mechanics, statistical models had always been considered to be less comprehensive than dynamic models. However, after Bohr proposed his thesis of completeness of quantum mechanics (as part of his argument with Einstein) and von Neumann instead presented his famous proof of impossibility of introducing hidden variables into quantum mechanics, the situation dramatically changed. At that moment it was believed that quantum mechanics was perfectly complete and its statistical description could never be substituted by deterministic predictions. From that perspective, the Bell model appeared to be a bright approach that, using the simplest hypothetical example, demonstrated the contrary. In fact, if hidden variables are non-observable and are uniformly distributed then there is no way to distinguish between deterministic and statistical models. In other words, a statistical model can be completed to a deterministic one by introducing hidden unobservable variables.

If we are to adopt a statistical quantum framework, then we can neither prove nor reject the existence of such variables. It is natural to call the phenomenon Gödel-type incompleteness and, in some sense, non-observability of hidden variables is the price to be paid in order for quantum theory to be consistent. If one could detect the hidden variables, then by properly selecting them, it would have been possible to derive results, contradicting with statistical predictions of quantum theory.

The famous Gödel's incompleteness theorem [4] states that if the theory is self-consistent and contains a limited number of axioms then it will always be possible to make a proposition that may be neither accepted nor rejected in the given framework. Of course, here we refer only to the analogy and not to the equality between the our example and the Gödel theorem.

The analogy marks up the hypothetical plausibility of introducing hidden variables allowing for transformation of statistical predictions into deterministic ones. Such a hypothesis can be neither rejected nor proved in the statistical framework.

The very formulation of the problem of Gödel -type incompleteness in classical probability theory is meaningless, as before the advent of quantum mechanics the statistical approach was a priori considered as inherently incomplete [2,5].

However, after von Neumann comprehensively described the statistical model of quantum mechanics using Hilbert space geometry [6], the quantum statistical science became as noble as Euclid geometry or arithmetic.

In this work we shall attempt to make two propositions that may seem mutually exclusive at first. The first one considers dynamic (Gödel type) incompleteness of quantum mechanics, while the second one actually points at statistical completeness. Thus, it is argued that the problem of completeness of quantum mechanics should be considered with regard to the interpretation of its formal mathematical apparatus.

If we aim to construct a dynamic deterministic theory, then quantum mechanics should be treated as an inherently incomplete formal model. Furthermore, quantum mechanics will be immediately rejected as soon as we discover the hidden variables that account for determinism of quantum theory. Such a possibility can not be neglected as demonstrated by the simple Bell example. Still, in 80 years from the birth of quantum mechanics there has not been a single serious candidate for the new theory, which of course does not imply that it may not appear in the future.

From the other perspective, if we may consider statistical interpretation of quantum mechanics, then the quantum theory appears to be perfectly complete (in the sense of completeness in quantum statistics, which is more comprehensive than standard probability theory). In particular, it appears that the attempts to introduce hidden variables to describe quantum statistics by classical probability distributions are necessarily fruitless.

### 1.6 Generalization of Bell's model. Dynamic chaos.

A complete theory, as noted by Bell in his work [1], has to be able to describe dynamics of hidden variables during measurement. However, Bell believes that the problem of measurement is extremely complicated for any theory (both with and without hidden variables) and for that reason he avoids a comprehensive analysis of the problem.

Note, nonetheless, that the model with hidden variables as proposed by Bell allows for a simple and natural generalization regarding the measurement problem. What is he most interesting aspect of the model is that analysis of dynamics of hidden variables during measurements would allow for generalization of quantum mechanics, which would asymptotically yield the same predictions.

Let $\lambda_0 = \sin^2(\theta/2) - 0.5$ be a point dividing the segment $-0.5 \le \lambda \le 0.5$ in a manner described above. Thus, the left side length equals to $\sin^2(\theta/2)$ (spin downwards), and the right side has the length of $\cos^2(\theta/2)$ (spin upwards). A uniformly filled segment $-0.5 \le \lambda < \lambda_0$ would correspond to the beam of particles deviated downwards, while the segment $\lambda_0 < \lambda \le 0.5$ - to the ones deviated upwards. If the hidden variable is to remain unchanged at the measurement, that fact would contradict with the predictions of quantum theory. In order to avoid it, the distribution of hidden variable in each of the two parts should yield the prior uniform distribution on $-0.5 \le \lambda \le 0.5$. The most natural way to fulfill the requirement is to implement a uniform extension that would transform each of the segments $-0.5 \le \lambda < \lambda_0$ and $\lambda_0 < \lambda \le 0.5$ into $-0.5 \le \lambda \le 0.5$.

Formally, the transformation is described as:

$$\lambda' = \begin{cases} \dfrac{\lambda + 0.5}{\lambda_0 + 0.5} - 0.5; & -0.5 \le \lambda < \lambda_0 \\ 0.5 - \dfrac{0.5 - \lambda}{0.5 - \lambda_0}; & \lambda_0 < \lambda \le 0.5 \end{cases}$$

The extension defines the desired hypothetical transformation of hidden variables during measurements.

It is evident that the transformation leads to chaotic dynamics in the system under consideration. The distance between any two initially close points grows exponentially (in geometrical progression after each measurement). Therefore, even a small error in the value of hidden variable $\lambda$ dramatically grows after a number of measurements, which makes impossible any deterministic prediction (one has to take the variable as uniformly distributed on $-0.5 \le \lambda \le 0.5$).

The system considered is perfectly analogous to the so-called Bernoulli shift [7-10]. Simple systems such as the Bernoulli shift, as demonstrated in the works of Prigogine and his followers, can simulate stochastic behavior of complex non-integrating dynamic systems [7].

A remarkable feature of such systems is that any initial distribution of hidden variables evolves to a uniform distribution over time.

Density dynamics of hidden variable distribution during measurements is described by the following recurrent equation (Perron- Frobenius transformation [7-10]):

$$\rho_{n+1}(\lambda) = (0.5+\lambda_0)\rho_n[(0.5+\lambda)(0.5+\lambda_0)-0.5] + (0.5-\lambda_0)\rho_n[0.5-(0.5-\lambda)(0.5-\lambda_0)] ,$$

where $\lambda_0 = \sin^2(\theta/2) - 0.5$

This equation describes the dynamics of hidden variable $\lambda$'s probability density during measurements (from the $n$-th to the $n+1$-th step).

An example of modeling of hidden variable distribution density dynamics is presented on Figure 2. It is evident that after a relatively small number of measurements the distribution already becomes almost uniform. A theory, where the initial distribution of hidden variables is not uniform is not in agreement with the predictions of quantum mechanics. However, asymptotically after a large number of measurements the results appear to be perfectly consistent. Thus, quantum mechanics may be considered as an asymptotical limit of a deterministic theory with hidden variables subject to dynamic chaos.

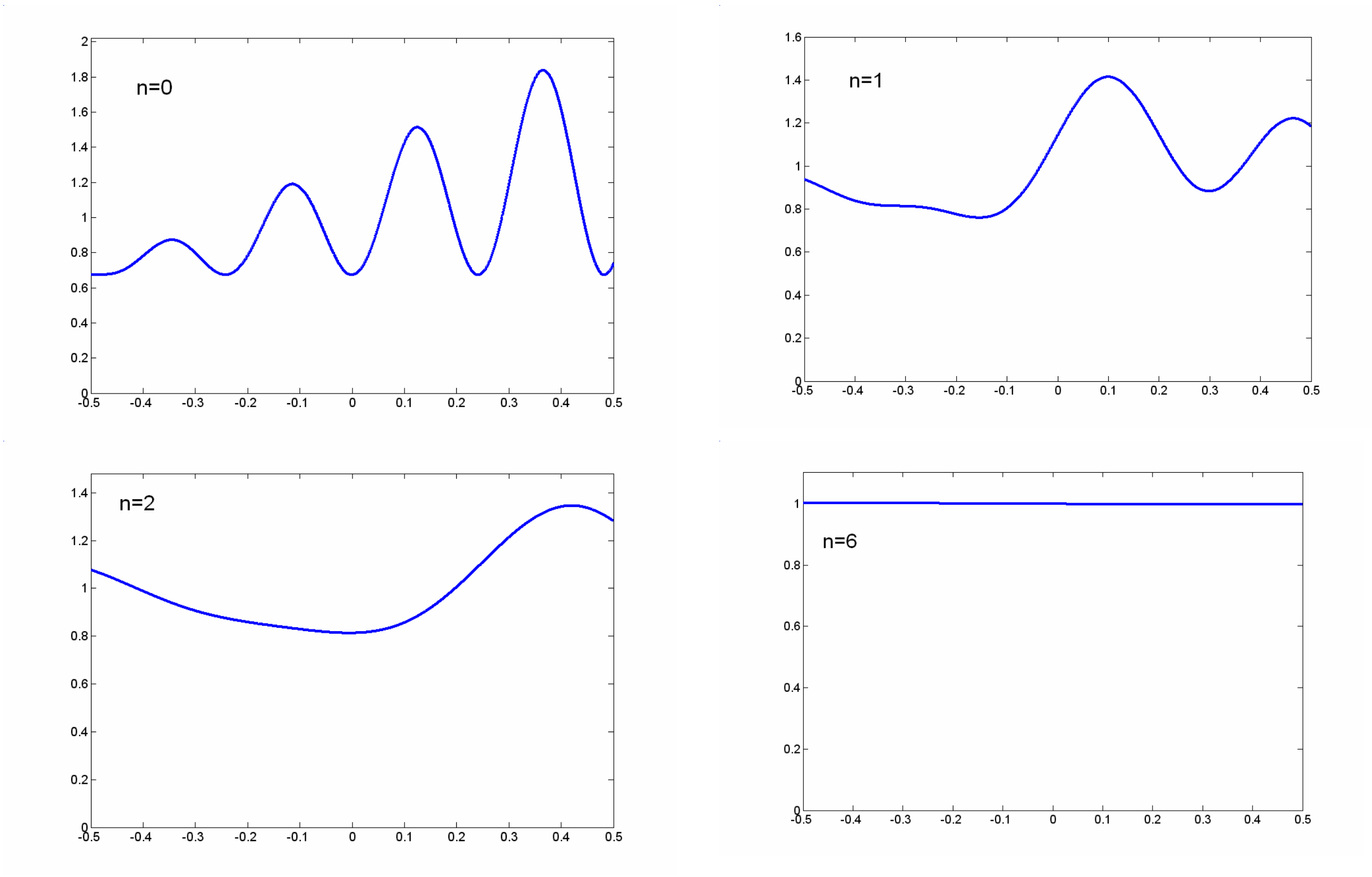

Figure 2. Example of hidden variable density dynamics.

Therefore, it is clear that description of dynamics of hidden variables is essential for an adequate description of the properties of a wave function reductions during measurements.

Such a problem does not arise when describing unitary evolution (then it is sufficient to suppose that the hidden variable does not change its value during unitary evolution)

In the framework of quantum informatics the model of a ½-spin particle is equivalent to a single qubit. The theory may also be easily generalized on arbitrary multi-cubit systems. Accordingly, instead on qubits one may consider arbitrary multi-level quantum systems (qudits) as well as systems with continuous spectrum.

The transformation discussed above corresponds to an instant extension. Let us consider an extension performed over time. Suppose that $\lambda$ corresponds to the origin time ($t = 0$), while $\lambda'$ is associated with infinitely large time ($t \to \infty$). Then we could use the following transform characteristic in our model:

$$\lambda(t) = \lambda + (\lambda' - \lambda)(1 - \exp(-t/\tau))$$

Here $\tau \to 0$ corresponds to quantum behavior (instant extension, "quantum leap" or "quantum jump"), and $\tau \to \infty$ stands for a classical urn scheme (constant (static) values of the hidden variable).

Given the framework, consider the following simple example. Let us have a beam of ½-spin particles, polarized along axis $z$ (initial spin upwards for all particles). Let us conduct consequent spin measurements: first along axis $x$ (measurement of $\sigma_x$), then along $z$ (measurement of $\sigma_z$). In a classical urn scheme all particles after the second measurement are bound to have spin upwards, as in the classical model measurement of $\sigma_x$ would not affect measurement of $\sigma_z$. The classical case corresponds to $t = 0$ on Fig.3. At the same time, $t \to \infty$ corresponds to the quantum behavior. The extension transformation during the measurement of $\sigma_x$ changed the hidden variable in the way that for the second measurement of $\sigma_z$ only half of the particles preserved spin upwards, while the other half became oriented downwards. (that is the result of the incompatibility between observables $\sigma_z$ and $\sigma_x$).

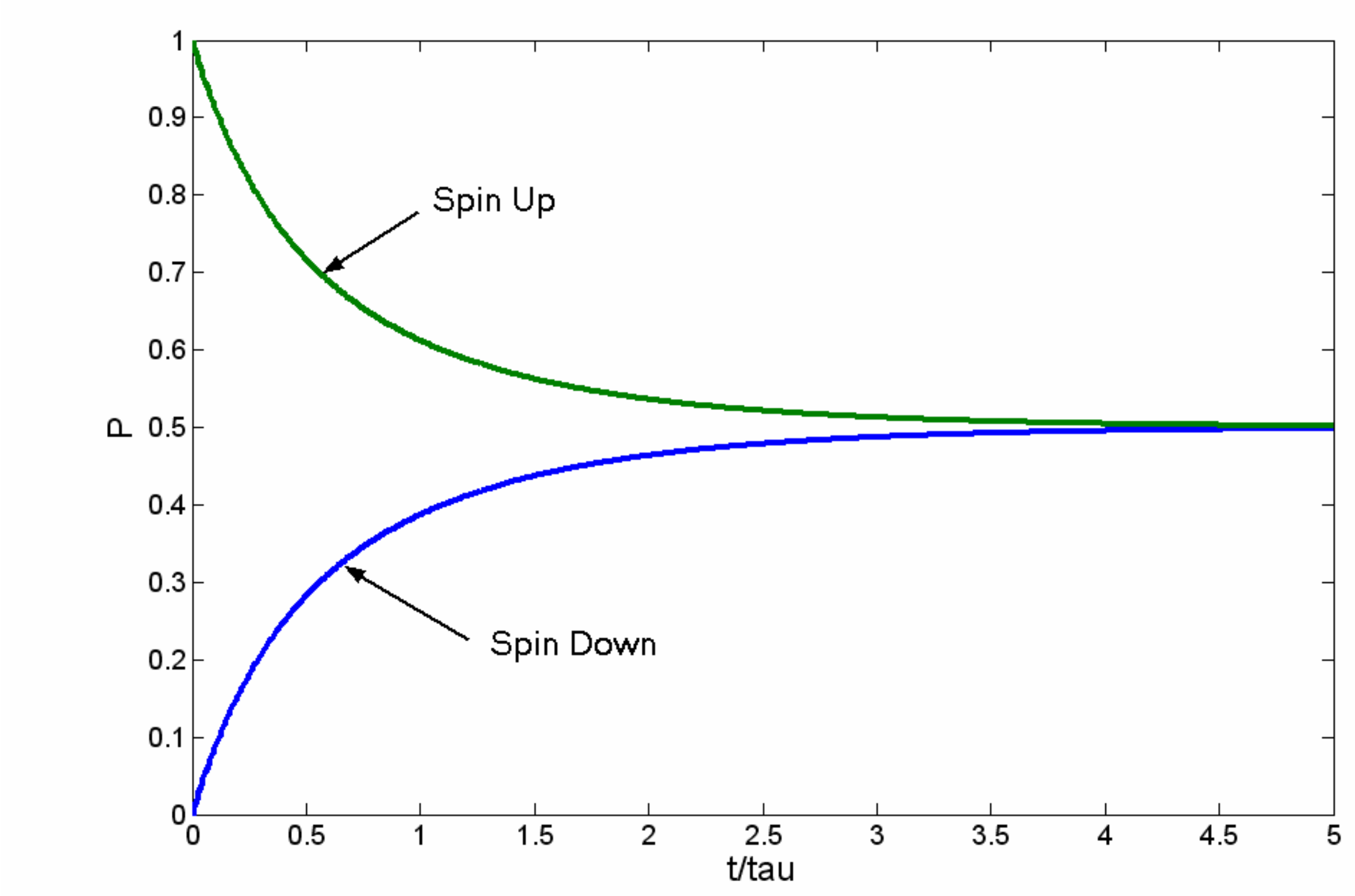

Fig.3. Spin measurement in the model corresponding to extension transformation that occurs gradually over time.

**1.7 Bell vs. von Neumann**

From Bell's perspective [1], there are significant drawbacks in von Neumann's proof of the famous theorem stating impossibility of introducing hidden variables to quantum mechanics. Actually, Bell heavily criticizes the approach in psychological perspective, which is further discussed in chapters 4.1 and 4.2. bell believes that the initial von Neumann's assumptions exclude the very possibility of introduction of hidden variables. Here we refer to the critique of condition II ([6], p. 234 of the Russian edition, p. 314 of the English edition). Let us explain the meaning of the condition.

In a theory with hidden variables one puts some classical random values into correspondence to quantum mechanical observables (Hermitian operators). It is well known that a linear combination (with real value coefficients) of two Hermitian operators (observables) is a Hermitian operator itself. Von Neumann's proposition was that a (hidden) classical variable, corresponding to the latter operator, was the same linear combination of two hidden variables of the former operators. For instance, if one associates random values $X$ и $Y$ with operators $\hat{X}$ и $\hat{Y}$ ($\hat{X} \to X$, $\hat{Y} \to Y$), then she should associate random value $aX + bY$ with operator $a\hat{X} + b\hat{Y}$. However, while this proposition seems natural, it is not exactly correct if operators $\hat{X}$ and $\hat{Y}$ are non-commuting (are incompatible). Consider the same ½-spin example with hidden variables proposed by Bell. Von Neumann's condition implies that the mean value of random value associated with operator $\alpha + \vec{\beta}\vec{\sigma}$ should be a linear combination of $\alpha$ and $\vec{\beta}$. This, however is not the case for a hypothetical Bell state with a given value of hidden parameter $\lambda$. The result of measurement for the state is determined, therefore there is no statistical variance and the observable value (mean value) is one of the eigenvalues $\alpha \pm |\vec{\beta}|$, which can not be represented as a linear combination of $\alpha$ and $\vec{\beta}$.

A hidden parameter $-0.5 \le \lambda \le 0.5$ determines the probability space, where the random hypothetical functions ($\hat{X} \to X$, $X = X(\lambda)$) are defined. It can be demonstrated by explicitly considering random values $X = X(\lambda)$ that the Bell's example violates the von Neumann's condition, as shown on the Figure.

Example. Let the spin of a particle be polarized along axis $Z$. Consider observable $\vec{\beta}\vec{\sigma}$, where the unit vector $\vec{\beta}$ lies on plane $(x,z)$. Let vector $\vec{\beta}_1$ for observable $\hat{X}_1 = \vec{\beta}_1\vec{\sigma}$ form angle $\theta_1 = \pi/6$ with axis $Z$, i.e. $\vec{\beta}_1 = \left(\frac{1}{2}, 0, \frac{\sqrt{3}}{2}\right)$. By analogy, angle $\theta_2 = \pi/3$ corresponds to observable $\hat{X}_2 = \vec{\beta}_2\vec{\sigma}$ i.e. $\vec{\beta}_2 = \left(\frac{\sqrt{3}}{2}, 0, \frac{1}{2}\right)$. Then observable $\hat{X}_3 = \vec{\beta}_3\vec{\sigma}$, for which vector $\vec{\beta}_3$ has angle $\theta_3 = \pi/4$ to axis $Z$, can be considered as a linear combination of observables $\hat{X}_1$ and $\hat{X}_2$. Indeed, as $\vec{\beta}_3 = \frac{\sqrt{2}}{1+\sqrt{3}}(\vec{\beta}_1 + \vec{\beta}_2) = \left(\frac{1}{\sqrt{2}}, 0, \frac{1}{\sqrt{2}}\right)$, then $\hat{X}_3 = \frac{\sqrt{2}}{1+\sqrt{3}}\left(\hat{X}_1 + \hat{X}_2\right)$ (see Figure).

On Figure 3, representations of random value $X_3$ in the frameworks of Bell and von Neumann are depicted. In Bell's model, in accordance with predictions of quantum mechanics, random parameter $X_3$ can take one of two possible values – either -1 or +1. At the same time, in von

Neumann's model, $X_3$ can take any of three values - $\frac{-2\sqrt{2}}{1+\sqrt{3}}$, 0 or $\frac{2\sqrt{2}}{1+\sqrt{3}}$, which clearly contradicts quantum mechanics. Note that as expected both models give correct mean values $M(X_3)=\frac{\sqrt{2}}{2}$.

There are two possible ways to resolve the contradiction between von Neumann's model and quantum mechanics. The first approach is to believe that the contradiction stands for inconsistency of the theory of hidden variables (this conclusion is made in the famous von Neumann theorem). Another possibility (following Bell) is to abandon the von Neumann theorem's condition, which states that a hypothetical hidden random value set into correspondence to two inconsistent observables is a sum of random values that correspond to each of the observables.

In the discussion that follows we will demonstrate that both approaches (which may be referred to as the direct and the indirect approaches) are reasonable.

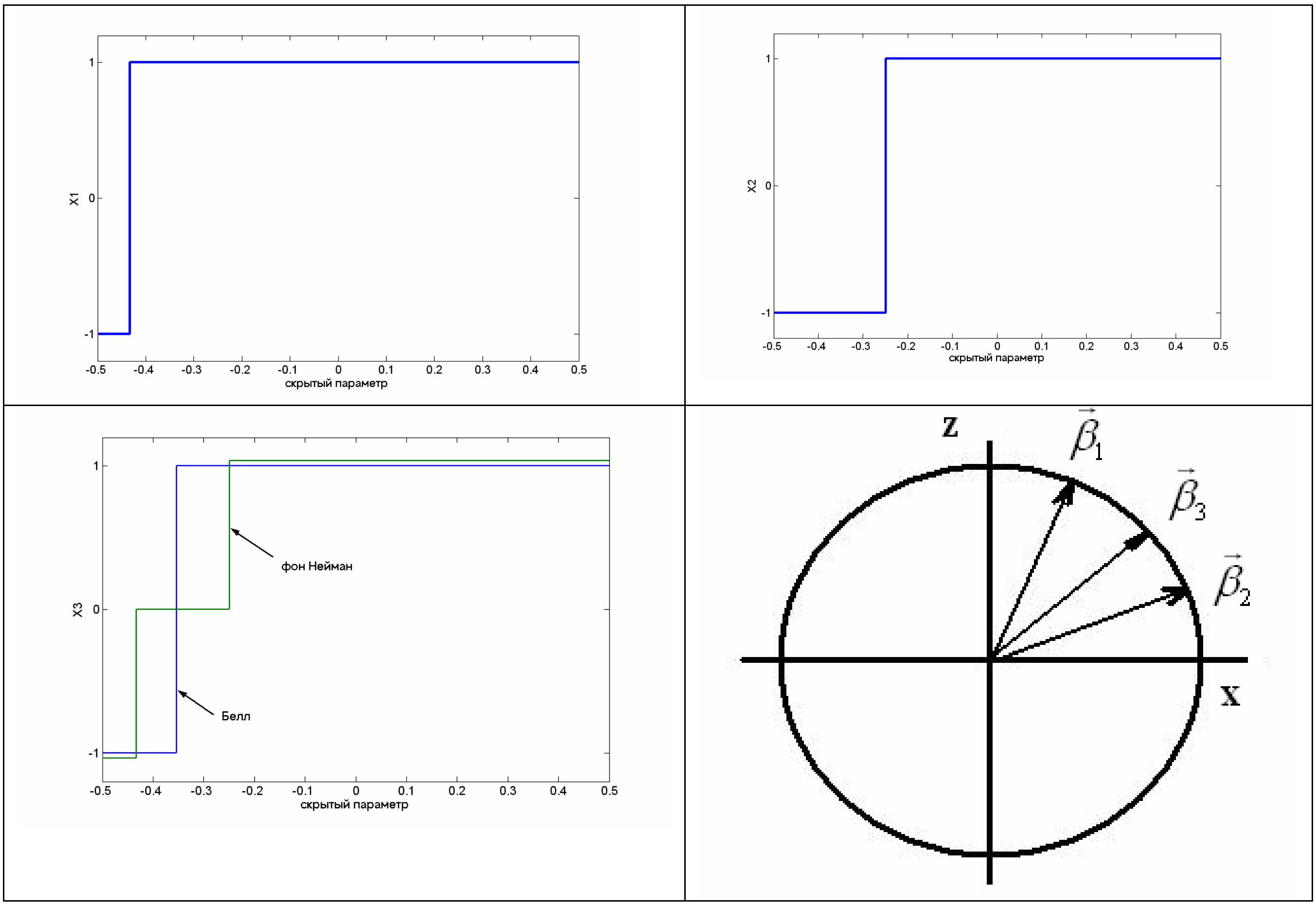

Figure 4. Comparison of representations of random values in Bell's and von Neumann's frameworks.

## 2. Hidden variables and impossibility of describing quantum statistics by classical probability theory.

### 2.1. Bell inequality as a paradox between statistics and quantum mechanics.

The problem of search for hidden variables is usually associated with the infamous Bell inequalities [11]. For the sake of convenience, we shall formulate the inequalities as a paradox (which will appear to be imaginary). This chapter is devoted to a discussion of this issue.

Consider the following quantum state which is sometimes referred to as a singlet Bell state:

$$|\psi\rangle = \frac{1}{\sqrt{2}}\left(|\uparrow\downarrow\rangle - |\downarrow\uparrow\rangle\right) \tag{1}$$

In the equation it is implied that the state consists of two spin particles. Alternatively, the state may be expressed as:

$$|\psi\rangle = \frac{1}{\sqrt{2}}\left(|01\rangle - |10\rangle\right) = \frac{1}{\sqrt{2}}\begin{pmatrix} 0 \\ 1 \\ -1 \\ 0 \end{pmatrix} \tag{2}$$

Here the Bell state is described as a two-qubit state.

We shall use operator $\vec{\sigma}$ as the spin observable (thus, we shall omit multiplier $\frac{\hbar}{2}$). Then, a measurement of the spin results either in $+1$, or $-1$.

If Alice gets $+1$ when measuring along axis $z$, then Bob will definitely get $-1$ for measurements along the same axis, and the other way round. This phenomenon does not depend on the choice of axis and it is due to the singleton property of the state. A singlet state is a two-particle state with the total spin equal to zero (therefore, the spin projection on any axis is also equal to zero).

Note that state $\frac{1}{\sqrt{2}}\left(|01\rangle + |10\rangle\right)$, which has a different operand ("plus" instead of "minus") from our state, also corresponds to a zero projection of the spin, but the total spin is equal to unity. A set of three states $|00\rangle$, $\frac{1}{\sqrt{2}}\left(|01\rangle + |10\rangle\right)$ and $|11\rangle$ forms a so-called triplet state. For a triplet, the total spin of two particles is equal to unity ($j = 1$), and there are three possible projections of the spin: $m = +1, 0, -1$.

One can easily show that the singlet state is invariant to the choice of quantization axis. Let $|0\rangle$ and $|1\rangle$ correspond to the $+1$ and $-1$ projections of the spin (operator $\sigma_z$) on some axis $\vec{n}$. States $|0'\rangle$ and $|1'\rangle$ correspond to another axis $\vec{n}'$. The former and the latter basis states are interconnected by a unitary transformation.

$$|0'\rangle = u_{00}|0\rangle + u_{01}|1\rangle$$

$$|1'\rangle = u_{10}|0\rangle + u_{11}|1\rangle$$

Here $U = \begin{pmatrix} u_{00} & u_{01} \\ u_{10} & u_{11} \end{pmatrix}$ is a unitary matrix.

Let its determinant be equal to unity:

$$u_{00}u_{11} - u_{10}u_{01} = 1$$

Then, the following relation holds true with necessity:

$$|0'1'\rangle - |1'0'\rangle = |01\rangle - |10\rangle \tag{3}$$

It implies that the singlet state is the same regardless of the quantization axis.

Note that the determinant of a unitary matrix may have phase multiplier, which is, however, insignificant for our consideration.

Now let us consider a procedure for singlet state measurement. Let Alice measure the projection of spin of her particle along axis $\vec{a}$, and Bob – projection of spin of his particle along axis $\vec{b}$.

Alice gets $+1$ with probability $\frac{1}{2}$ and $-1$ with probability $\frac{1}{2}$. After that, the state reduces in a way that for measurements along the same axis $\vec{a}$ Bob gets $-1$, if Alice measures $+1$ and vice versa. If Bob conducts his measurement along some other axis $\vec{b}$, oriented at angle $\theta$ to the Alice's axis, then in accordance with the results presented in chapter 1, one gets the following probabilities:

$$P_{AB}(+1,-1) = \frac{1}{2}\cos^2\left(\frac{\theta}{2}\right) \quad (4)$$

$$P_{AB}(+1,+1) = \frac{1}{2}\sin^2\left(\frac{\theta}{2}\right) \quad (5)$$

$$P_{AB}(-1,+1) = \frac{1}{2}\cos^2\left(\frac{\theta}{2}\right) \quad (6)$$

$$P_{AB}(-1,-1) = \frac{1}{2}\sin^2\left(\frac{\theta}{2}\right) \quad (7)$$

Here $P_{AB}(+1,-1)$ says that Alice gets $+1$, while Bob gets $-1$ and so on.

It is clear that marginal distributions describing Bob and Alice's individual probabilities are:

$$P_A(+1) = P_A(-1) = \frac{1}{2} \qquad P_B(+1) = P_B(-1) = \frac{1}{2} \quad (8)$$

Mathematical expectations of the distributions are equal to zero, while the variances are equal to unity.

Let $X$ and $Y$ be random values registered by Alice and Bob. The correlation coefficient of the values is:

$$R_{AB} = M(XY) = -\cos(\theta) = -\vec{a}\vec{b} \quad (9)$$

It appears that quantum correlations that are observed in Bell's singlet state reject the approach as the correlations may not be simulated by any classical scheme with hidden variables.

Let us consider the so-called Bell-Clauser-Horne-Shimony-Holt inequality [12].

Let $X_1$, $X_2$, $Y_1$, $Y_2$ - be arbitrary real numbers not greater than 1 in absolute value. $|X_j| \le 1$, $|Y_j| \le 1$, $j = 1,2$

The inequality, which is elementary derived, has the following representation:

$$-2 \le X_1Y_1 + X_1Y_2 + X_2Y_1 - X_2Y_2 \le 2 \quad (10)$$

To prove it, let all the variables be non-negative and $Y_1 \ge Y_2$. Then

$$X_1Y_1 + X_1Y_2 + X_2Y_1 - X_2Y_2 = X_1(Y_1 + Y_2) + X_2(Y_1 - Y_2) \leq$$
$$\max(X_1, X_2)(Y_1 + Y_2 + Y_1 - Y_2) = 2Y_1 \max(X_1, X_2) \leq 2$$

All other possibilities are treated by analogy, which proves the inequality.

Now let $X_1$, $X_2$, $Y_1$, $Y_2$ - be real random values, which satisfy the inequalities.

It can be easily shown that the inequalities that are valid for some random values remain valid for the corresponding mean values as well.

Then taking the average of the Bell-Clauser-Horne-Shimony-Holt inequality we get the following:

$$\left|M(X_1Y_1) + M(X_1Y_2) + M(X_2Y_1) - M(X_2Y_2)\right| \leq 2 \quad (11)$$

The result is usually referred to as the Bell inequality. It may appear that the inequality derived from simple and intuitive assumptions should remain valid for arbitrary statistical system. However, it is not true. As we will see, the inequality is violated during Bell's singlet state measurements. This contradiction between the seemingly general statistical result and predictions of quantum mechanics is the heart of the paradox with the Bell's inequalities.

In fact, let us choose the measurements' directions on one plane as shown on Fig. 5. [12]:

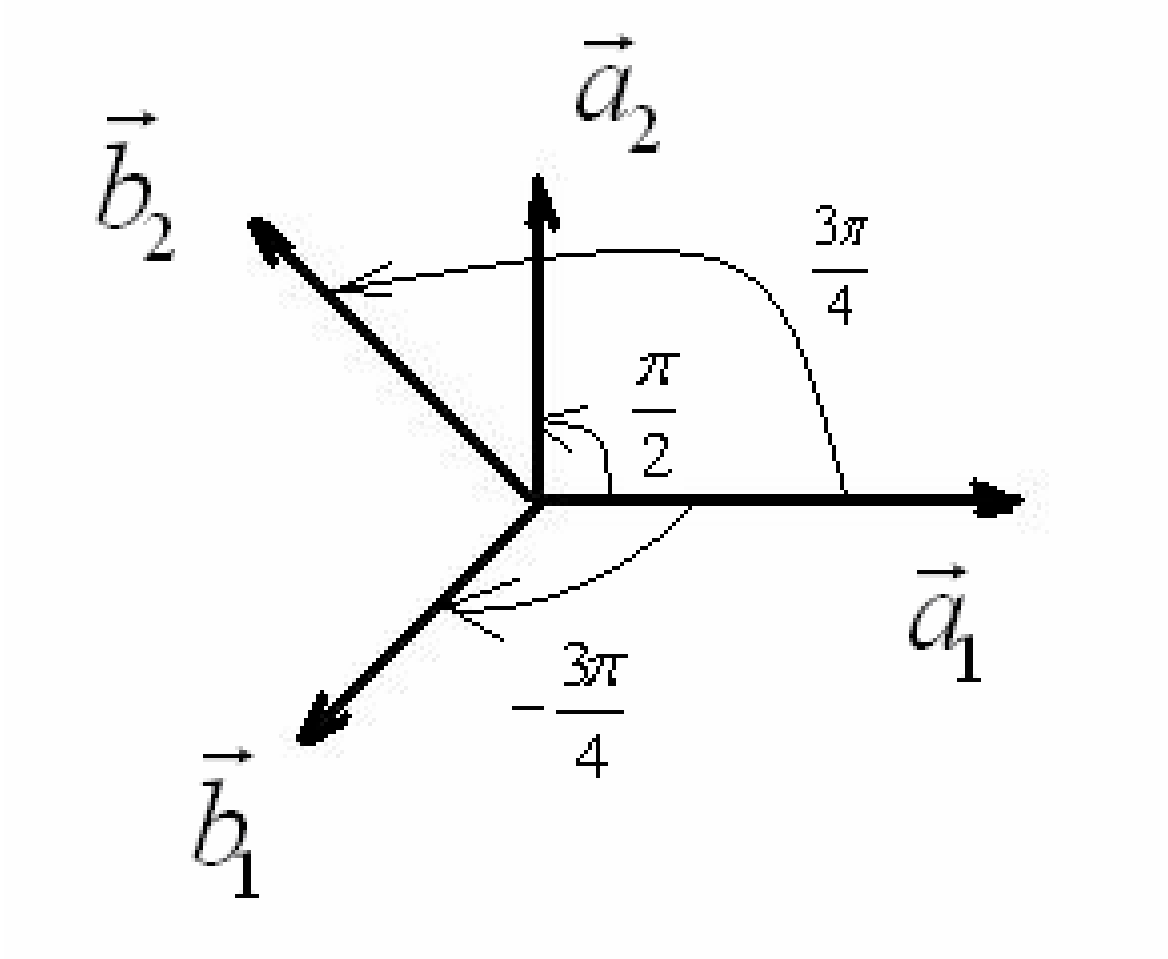

Fig. 5. Geometry of singleton state measurement scheme.

$$\varphi = 0 \text{ for } \vec{a}_1,\ \varphi = \frac{\pi}{2} \text{ for } \vec{a}_2,\ \varphi = -\frac{3\pi}{4} \text{ for } \vec{b}_1,\ \varphi = \frac{3\pi}{4} \text{ for } \vec{b}_2. \quad (12)$$

Then:

$$M(X_1Y_1) = M(X_1Y_2) = M(X_2Y_1) = -\cos\left(\frac{3\pi}{4}\right) = \frac{\sqrt{2}}{2} \quad (13)$$

$$M(X_2Y_2) = -\cos\left(\frac{\pi}{4}\right) = -\frac{\sqrt{2}}{2} \quad (14)$$

As the result:

$$M(X_1Y_1) + M(X_1Y_2) + M(X_2Y_1) - M(X_2Y_2) = 2\sqrt{2} > 2 \quad (15)$$

Therefore, the Bell's inequality is violated.

Let us clarify the statistical meaning of the Bell's inequality and the fact of its violation. During the averaging procedure of the Bell-Clauser-Horne-Shimony-Holt inequality (calculations of $M(X_1Y_1), M(X_1Y_2), M(X_2Y_1)$ and $M(X_2Y_2)$), it was implicitly stated that there existed joint distribution of random variables $P(X_1, Y_1, X_2, Y_2)$ (the total of 16 probabilities). In reality, such distribution does not exist in the example above. In other words, it was impossible to take 16 non-negative values (probabilities) to explain all correlations. Some of the probabilities would have to be negative, and therefore would not be real probabilities. This question is discussed in detail in the following chapter.

**2.2. Reconstruction of hypothetical joint distribution in Bell's problem. Singular value decomposition (svd) approach.**

Consider a hypothetical joint distribution $P(X_1, X_2, Y_1, Y_2)$. There are 16 values that it can take, which we will denote according to Table 1.

| | | | |
|---|---|---|---|
| $P(+1,+1,+1,+1)=P_1$ | $P(+1,+1,+1,-1)=P_2$ | $P(+1,+1,-1,+1)=P_3$ | $P(+1,+1,-1,-1)=P_4$ |
| $P(+1,-1,+1,+1)=P_5$ | $P(+1,-1,+1,-1)=P_6$ | $P(+1,-1,-1,+1)=P_7$ | $P(+1,-1,-1,-1)=P_8$ |
| $P(-1,+1,+1,+1)=P_9$ | $P(-1,+1,+1,-1)=P_{10}$ | $P(-1,+1,-1,+1)=P_{11}$ | $P(-1,+1,-1,-1)=P_{12}$ |
| $P(-1,-1,+1,+1)=P_{13}$ | $P(-1,-1,+1,-1)=P_{14}$ | $P(-1,-1,-1,+1)=P_{15}$ | $P(-1,-1,-1,-1)=P_{16}$ |

These probabilities are hypothetical (fiction like) as the 4 observables are incompatible. We shall use the probabilities as a hypothetical basis for 2-dimensional distributions which can be measured in practice, i.e. $P(X_1, Y_1)$, $P(X_1, Y_2)$, $P(X_2, Y_1)$, $P(X_2, Y_2)$.

Based on eq (4)-(7) we shall derive relationships interconnecting the measurable 2D-distributions and the hypothetical joint distribution

1. Measurement of distribution $P(X_1, Y_1)$: $\theta = \theta_{11} = 2\pi - \frac{3\pi}{4} = \frac{5\pi}{4}$

$$P_1 + P_2 + P_5 + P_6 = \frac{1}{2}\sin^2\left(\frac{\theta_{11}}{2}\right); \qquad P_3 + P_4 + P_7 + P_8 = \frac{1}{2}\cos^2\left(\frac{\theta_{11}}{2}\right);$$

$$P_9 + P_{10} + P_{13} + P_{14} = \frac{1}{2}\cos^2\left(\frac{\theta_{11}}{2}\right); \qquad P_{11} + P_{12} + P_{15} + P_{16} = \frac{1}{2}\sin^2\left(\frac{\theta_{11}}{2}\right)$$

2. Measurement of distribution $P(X_1, Y_2)$: $\theta = \theta_{12} = \frac{\pi}{2} + \frac{\pi}{4} = \frac{3\pi}{4}$

$$P_1 + P_3 + P_5 + P_7 = \frac{1}{2}\sin^2\left(\frac{\theta_{12}}{2}\right); \qquad P_2 + P_4 + P_6 + P_8 = \frac{1}{2}\cos^2\left(\frac{\theta_{12}}{2}\right);$$

$$P_9 + P_{11} + P_{13} + P_{15} = \frac{1}{2}\cos^2\left(\frac{\theta_{12}}{2}\right); \qquad P_{10} + P_{12} + P_{14} + P_{16} = \frac{1}{2}\sin^2\left(\frac{\theta_{12}}{2}\right)$$

3. Measurement of distribution $P(X_2, Y_1)$: $\theta = \theta_{21} = \frac{3\pi}{4}$

$$P_1 + P_2 + P_9 + P_{10} = \frac{1}{2}\sin^2\left(\frac{\theta_{21}}{2}\right); \qquad P_3 + P_4 + P_{11} + P_{12} = \frac{1}{2}\cos^2\left(\frac{\theta_{21}}{2}\right);$$

$$P_5 + P_6 + P_{13} + P_{14} = \frac{1}{2}\cos^2\left(\frac{\theta_{21}}{2}\right); \qquad P_7 + P_8 + P_{15} + P_{16} = \frac{1}{2}\sin^2\left(\frac{\theta_{21}}{2}\right)$$

4. Measurement of distribution $P(X_2, Y_2)$: $\theta = \theta_{22} = \frac{\pi}{4}$

$$P_1 + P_3 + P_9 + P_{11} = \frac{1}{2}\sin^2\left(\frac{\theta_{22}}{2}\right); \qquad P_2 + P_4 + P_{10} + P_{12} = \frac{1}{2}\cos^2\left(\frac{\theta_{22}}{2}\right);$$

$$P_5 + P_7 + P_{13} + P_{15} = \frac{1}{2}\cos^2\left(\frac{\theta_{22}}{2}\right) \; ; \qquad P_6 + P_8 + P_{14} + P_{16} = \frac{1}{2}\sin^2\left(\frac{\theta_{22}}{2}\right)$$

The relationships presented above set a system of 16 equations with 16 variables (some of them are dependent, which will be discussed below)

Let us represent the system in the matrix form:

$Ap = b$, where $p$ - is a column of 16 unknown probabilities, presented above, and $A$ - is a matrix $16 \times 16$, consisting of zeros and ones.

$$A = \begin{pmatrix}
1 & 1 & 0 & 0 & 1 & 1 & 0 & 0 & 0 & 0 & 0 & 0 & 0 & 0 & 0 & 0 \\
0 & 0 & 1 & 1 & 0 & 0 & 1 & 1 & 0 & 0 & 0 & 0 & 0 & 0 & 0 & 0 \\
0 & 0 & 0 & 0 & 0 & 0 & 0 & 0 & 1 & 1 & 0 & 0 & 1 & 1 & 0 & 0 \\
0 & 0 & 0 & 0 & 0 & 0 & 0 & 0 & 0 & 0 & 1 & 1 & 0 & 0 & 1 & 1 \\
1 & 0 & 1 & 0 & 1 & 0 & 1 & 0 & 0 & 0 & 0 & 0 & 0 & 0 & 0 & 0 \\
0 & 1 & 0 & 1 & 0 & 1 & 0 & 1 & 0 & 0 & 0 & 0 & 0 & 0 & 0 & 0 \\
0 & 0 & 0 & 0 & 0 & 0 & 0 & 0 & 1 & 0 & 1 & 0 & 1 & 0 & 1 & 0 \\
0 & 0 & 0 & 0 & 0 & 0 & 0 & 0 & 0 & 1 & 0 & 1 & 0 & 1 & 0 & 1 \\
1 & 1 & 0 & 0 & 0 & 0 & 0 & 0 & 1 & 1 & 0 & 0 & 0 & 0 & 0 & 0 \\
0 & 0 & 1 & 1 & 0 & 0 & 0 & 0 & 0 & 0 & 1 & 1 & 0 & 0 & 0 & 0 \\
0 & 0 & 0 & 0 & 1 & 1 & 0 & 0 & 0 & 0 & 0 & 0 & 1 & 1 & 0 & 0 \\
0 & 0 & 0 & 0 & 0 & 0 & 1 & 1 & 0 & 0 & 0 & 0 & 0 & 0 & 1 & 1 \\
1 & 0 & 1 & 0 & 0 & 0 & 0 & 0 & 1 & 0 & 1 & 0 & 0 & 0 & 0 & 0 \\
0 & 1 & 0 & 1 & 0 & 0 & 0 & 0 & 0 & 1 & 0 & 1 & 0 & 0 & 0 & 0 \\
0 & 0 & 0 & 0 & 1 & 0 & 1 & 0 & 0 & 0 & 0 & 0 & 1 & 0 & 1 & 0 \\
0 & 0 & 0 & 0 & 0 & 1 & 0 & 1 & 0 & 0 & 0 & 0 & 0 & 1 & 0 & 1
\end{pmatrix}$$

$b$ - is the right side of the equation (column of 16 numbers)

It appears that only 9 out of the 16 equations are independent, thus the system has an infinitely large number of solutions. Still, any of the solutions should include negative numbers (negative probabilities). Obviously, physical probabilities may not take negative values and, therefore, the initial joint probability distribution does not exist.

Consider the results of numerical analysis of the system in more detail.

Let us apply so-called svd-decomposition (singular value decomposition) to matrix A. It allows us to present the matrix in the following form:

$A = USV^+$,

Where $U$ and $V$ - are unitary (orthogonal) matrices and $S$ - is a diagonal non-negatively defined matrix (diagonal elements of the matrix are called singular values). Based on the rank of the system, only 9 out of the 16 elements are non-zero.

Let us introduce a new variable (factor)

$f = V^+ p$ .

Then the system may be presented as:

$Sf = U^+ b$

This system is elementary as matrix $S$ is diagonal.

The 7 last diagonal elements of matrix $S$ and accordingly the 7 latest elements of column $U^+ b$ are inherently equal to zero. Thus, the system is self-consistent, while it has an infinitely large number of solutions. This is because, the latter 7 equations take the form $0f_i = 0 \quad i = 10,...,16$, and therefore for $i = 10,...,16$ factors $f_i$- are arbitrary real numbers. Solution of the first 9 equations is obviously $f_i = (U^+ b)_i / S_i \quad i = 1,...,9$, where $(U^+ b)_i$ is the $i$-th element of column $U^+ b$, and $S_i$ is the $i$-th diagonal element of matrix $S$. All possible vector-columns $f$ are derived in the manner. All possible solution-columns are derived as:

$$p = Vf$$

Among all the possible solutions there is a special one. It can be derived putting the 7 arbitrary values $f_i$ equal to zero: $f_i = 0 \quad i = 10,...,16$. It appears that the solution has a special property that distinguishes it from other possible solutions: it minimizes the sum of squares of probabilities and therefore the probability variance:

$$\sum_{i=1}^{16} p_i^2 = \min, \quad D(p) = \min$$

Note that the sum of probabilities according to the normalization condition is fixed and is equal to unity

$$\sum_{i=1}^{16} p_i = 1$$

According to applied mathematics, this distinguished solution may be referred to as the regularized solution. The right side of the equation ( column $b$ ) defines probabilities that may be really experimentally measured. Inevitable errors in the experiment result in a modification of the equations considered above. Equations $0f_i = 0 \quad i = 10,...,16$, will be substituted by $0f_i = \varepsilon_i \quad i = 10,...,16$. It means that instead of equations that have an infinite number of solutions we will get equations which have no solutions at all. Regularization implies that we identify the contradiction as being error-caused, instead of being caused by the nature of the problem, and therefore we shall omit $\varepsilon_i$ on the right side and set $f_i = 0 \quad i = 10,...,16$.

The column, which sets the regularized solution consists of positive elements $(1+\sqrt{2})/16$ and negative elements $(1-\sqrt{2})/16$. It has the following form:

$$p = \begin{pmatrix} (1+\sqrt{2})/16 \\ (1+\sqrt{2})/16 \\ (1-\sqrt{2})/16 \\ (1-\sqrt{2})/16 \\ (1+\sqrt{2})/16 \\ (1-\sqrt{2})/16 \\ (1+\sqrt{2})/16 \\ (1-\sqrt{2})/16 \\ (1-\sqrt{2})/16 \\ (1+\sqrt{2})/16 \\ (1-\sqrt{2})/16 \\ (1+\sqrt{2})/16 \\ (1-\sqrt{2})/16 \\ (1-\sqrt{2})/16 \\ (1+\sqrt{2})/16 \\ (1+\sqrt{2})/16 \end{pmatrix}$$

This result follows directly from the unitary (orthogonality) property of the transformation defined by matrix $V$. The sum of squares of elements is invariant in the transformation, therefore:

$$\sum_{i=1}^{16} p_i^2 = \sum_{i=1}^{16} f_i^2$$

The sum of squares is at its minimum, when each of the seven elements $f_i$ which can be set arbitrarily has zero value. This case corresponds to the regularized solution presented above..

There is another remarkable property of any solution of the system of equations, which is equivalent to Bell's inequalities.

Let us consider this property and introduce the following vector-row of 16 elements, each of which is equal either to +1 or -1.

$$a = (1 \ \ 1 \ \ -1 \ \ -1 \ \ 1 \ \ -1 \ \ 1 \ \ -1 \ \ -1 \ \ 1 \ \ -1 \ \ 1 \ \ -1 \ \ -1 \ \ 1 \ \ 1)$$

The value of each element of the vector-row $a$ is equal to the value of the corresponding algebraic expression $0.5(X_1Y_1 + X_1Y_2 + X_2Y_1 - X_2Y_2)$

It appears that the product of vector-row $a$ and some arbitrary solution $p$ (vector-column) is invariant:

$$ap = \sqrt{2}$$

Let us call it Bell-type invariant and explain its meaning in the framework of linear algebra. It is clear that $ap = (aV)f$. Moreover, the 7 last columns in matrix $V$ are orthogonal to vector $a$, and, therefore, the 7 last elements in vector-row $(aV)$ equal zero. Consequently, the result will not depend on the values of the 7 last elements of vector-column $f$, which could therefore be chosen in an arbitrary manner.

The properties discussed in this chapter are obviously the same logical arguments from chapter 2.1 presented in the framework of linear algebra.

For any values of physical probabilities (which have to be non-negative) condition $|ap| \leq 1$ (which is Bell's inequality) should be satisfied. To obtain $ap = \sqrt{2}$ one needs to impose negative values to some of the probabilities. It is well known that $\sqrt{2}$ is the level that demonstrates how many times the Bell's inequality is violated in the singlet state. Therefore, our analysis discovered the following fact: Bell's inequalities are nothing but the non-negativity restriction on the values of probabilities in a distribution. The Bell's invariant demonstrates that the desired probability distribution does not exist (there is an infinitely large number of solutions of the considered system of equations, which however inevitably contain non-physical negative probabilities)

Note that the same interpretation of Bell's inequality in the sense of existence of negative probabilities has already been discussed by Sudarshan and, Rothman [13,14]. Our analysis completes the discussion in some ways: analysis by means of svd-decomposition has been conducted, which enabled us to present all possible solutions in a compact form; a regularized solution corresponding to minimal variance has been obtained; an invariant, which explicitly demonstrated the nature of Bell's inequalities' violation, has been presented and analyzed.

### 2.3. Bell's inequality and dynamic hidden variables.

The ½-spin model (qubit) with hidden variables, described in chapter 1, can be generalized for arbitrary quantum systems. Let us consider a two-qubit model, for example. For each qubit we shall associate a random value uniformly distributed on [-0.5 0.5] ($-0.5 \leq \lambda_1 \leq 0.5$, $-0.5 \leq \lambda_2 \leq 0.5$). Then the space of hidden variables would be a unitary square. Now let us consider singlet state measurements by means of the hidden variables theory.

The results of the measurements are presented on Fig.6. The measurement protocol divides the unitary square in 4 pieces, with the squares of the pieces equal to the corresponding probabilities (results of measurements of $X_i$ and $Y_j$, $i,j = 1,2$ are presented by signs on the figure). Note that measurements of pairs $(X_1Y_1),(X_1Y_2),(X_2Y_1)$ are graphically described by the same picture (to the left) this is because in all of the cases the angle between axes directions is the same ($\frac{3\pi}{4}$) and measurements results are therefore the same too.

In the fourth pair $(X_2Y_2)$ the actual angle is equal to $\frac{\pi}{4}$ and, therefore, the probabilities of positive and negative projections switch to one another.

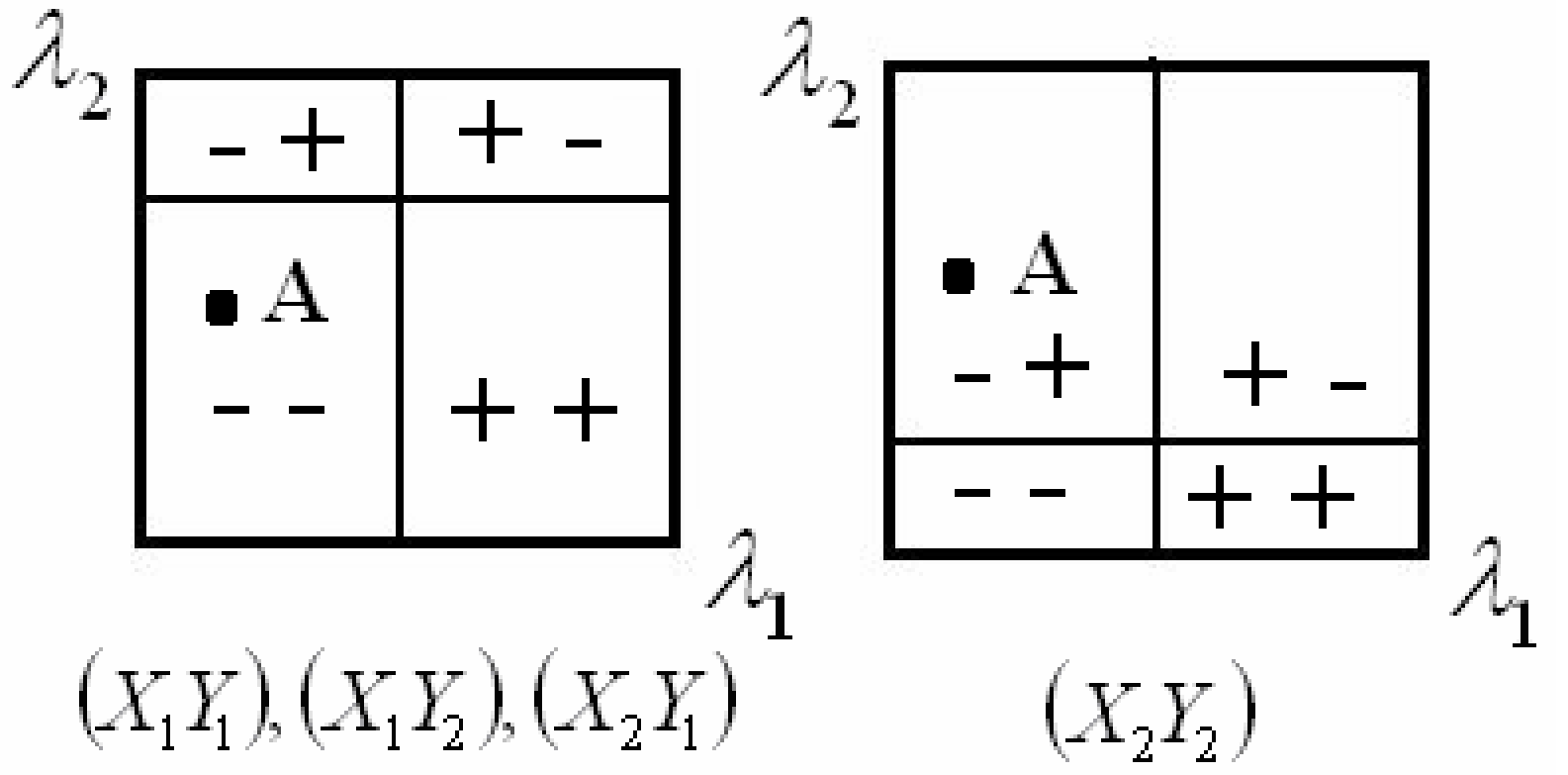

Fig. 6 Hidden variables in a 2-qubit singlet state measurement scheme.

Hypothetical hidden variables transform a statistical experiment into a deterministic one. Results of the experiment are in perfect agreement with predictions of quantum mechanics (in particular, Bell's inequalities are violated). Here we again refer to Gödel -type incompleteness as we did in the case of one qubit, because if hidden variables are uniformly distributed and are not observable then one cannot neither prove nor reject their existence.

Note that the analogy with the one-qubit model is more profound, as we may associate the transformation of extension with the procedure of measurement. Similar to the one-qubit case, any arbitrary non-uniform distribution of hidden variables will transform into a uniform one in its asymptotic limit (after a large number of measurements) .

Figure 6 geometrically illustrates the absence of a joint distribution of probabilities $P(X_1,X_2,Y_1,Y_2)$, which is crucial for explaining the violations of Bell's inequalities. For example, point A on the figure corresponds to $Y_2=-1$ for measurements in the pair $(X_1Y_2)$ and at the same time to $Y_2=+1$ for measurements in pair $(X_2Y_2)$. This (the differences in the left-hand and the right-hand figures) implies that the square of hidden variables may not be considered as the elementary probability space for a 4D distribution $(X_1,Y_1,X_2,Y_2)$: There is no unambiguous correspondence between the points in the square and hypothetical 4D measurements, because the value of $Y_2$ depends on the value it is measured with ( either $X_1$ or $X_2$). Such kinds of models are generally called contextual models, because the values of arbitrary variables are dependent on the context of the measurement (measurement condition)

**2.4. Incompatibility of quantum statistics with classical urn scheme: direct proof and indirect evidence.**

The impossibility of describing the results of quantum measurements by means of classical joint probability distributions has both direct and indirect aspects.

The direct aspect of the problem is evident from the following well-known example of a ½-spin particle. Let us consequently measure observables $\sigma_z$ and $\sigma_x$, associated with momentum projections on axes $Z$ and $X$ . After the first measurement we shall separate the particles that deviated upwards in the Stern-Gerlach device (which corresponded to $\sigma_z=+1$). Next, we shall measure observable $\sigma_x$ of the resulting beam with $\sigma_z=+1$ and we shall omit all the observables except $\sigma_x=+1$. In the classical framework, the particles in the final beam have both $\sigma_z=+1$ and $\sigma_x=+1$. However, surprisingly enough, this conclusion is not supported by the experiment! If we conduct measurements of $\sigma_z$ in the final beam again, then half of the particles would have $\sigma_z=-1$, despite the fact that originally only the $\sigma_z = +1$ particles were present. Thus, the results of the second measurement modify the result of the first measurement.

In our discussion we closely follow the one by von Neumann ([6], p. 300 -305) with the only exception that von Neumann refers to some abstract observables that can take two values instead of the spin operators. Von Neumann concludes that "We have no method ... to penetrate to those homogeneous ensembles which no longer have dispersion" ([6], p. 305)

The incompatibility of observables such as $\sigma_z$ and $\sigma_x$ is mathematically represented by means of non-commuting operators. Non-commuting operators have no common eigenvectors, and therefore there is no state, where both observables would have certain values. In quantum mechanics, incompatible observables result in such fundamental notions as the complementarity principle and the uncertainty relations [15-17].

Considering of the incompatible observables results in a violation of the known axiom of composite random values in classical probability theory [18]. According to Cramer [19], we may formulate the axiom as follows: "If $\xi_1,...,\xi_n$ are random values with dimensions $k_1,...,k_n$ correspondingly then each composite value $(\xi_1,...,\xi_n)$ too is a random value (of dimension $k_1+...+k_n$)". However, the axiom is violated in quantum mechanics because an object composed of mutually complementary random values is no longer a random value but it rather corresponds to a more general notion of a quantum state. A quantum state may be considered as a natural generalization of the notion of statistical distribution. According to the aforementioned reasons, a quantum state may not be described by a single statistical distribution but rather by a set of mutually-complementary distributions.

Von Neumann extends the argument of impossibility of simultaneous measurement of non-commuting observables. He argues that while it is true that if observables $\hat{X}$ and $\hat{Y}$ are non-commuting then their joint distribution $P(X,Y)$ does not make any sense (according to experimental and theoretical results in quantum mechanics). Still, as he extends his argument, let us imagine another observable $\hat{Z}$ that commutes both with $\hat{X}$ and $\hat{Y}$. Then we may describe distributions $P(X,Z)$ and $P(Y,Z)$. Then, observables $\hat{X}$ and $\hat{Y}$ become connected by means of observable $\hat{Z}$. Then the following question arises: would it be possible to consider the joint distribution $P(X,Y,Z)$ as some assisting (fiction) object? Doing so, we could derive the real experimentally measurable distribution $P_{xz}(X,Z)$ by summing the hypothetic distribution by variable $\hat{Y}$, i.e. $P_{xz}(X,Z)=\sum_y P(X,Y,Z)$ (distribution $P_{xz}(X,Z)$ is called marginal distribution with regards to a more complete distribution $P(X,Y,Z)$). The hypothesis of existence of such distribution $P(X,Y,Z)$ implies that observable $\hat{Y}$ is hidden when considering pair ($\hat{X}$, $\hat{Z}$), while $\hat{X}$ is thought to be hidden with regards to ($\hat{Y}$, $\hat{Z}$).

The sample space for hypothetic distribution $P(X,Y,Z)$ is a set of points in 3D space $(X,Y,Z)$. Each of the points is a hypothetical uniform ensemble with zero variance, where all considered observables ($\hat{X}$, $\hat{Y}$ и $\hat{Z}$) have defined values. Then, in von Neumann's words, the problem of hidden variables is derived to "trying to support the fiction view" of existence of zero-variance ensembles, and, therefore, the view of existence of a joint distribution for, generally speaking, inconsistent observables.

Note that the considered formulation of implicit conjunction of inconsistent observables can only be made in Hilbert spaces of dimensions of three and higher. In fact, in a 2D Hilbert space the following simple statement holds true (it can be described as relation transitivity of commutability of operators or consistency of observables): if operators $\hat{X}$ and $\hat{Y}$ are commuting with operator $\hat{Z}$, then they also commute with one another.

As a result, in a 2D case, the whole problem of implicit conjunction of inconsistent observables loses its gist. In a Hilbert space of dimension equal to three or higher the relation of consistency of observables is not transitive. In that case, the problem of implicit conjunction of observables has some sense.

Taking into account the facts presented above, let us consider the main Bell's result (his famous inequality). Consider the following problem: if (X1,X2) and (Y1,Y2) are inconsistent observables, while (X1,Y1), (X1,Y2), (X2,Y1) and (X2,Y2) are consistent, then does the probability distribution P(X1,X2,Y1,Y2) exist? Given this direct formulation, one may immediately give a negative answer to the question. If the distribution P(X1,X2,Y1,Y2) did exist, then the marginal distribution P(X1,X2) would also exist, which would imply that there would be a state with the values X1 and X2 simultaneously defined. However, the latter is not possible as operators X1 and X2 do not commute with each other. Therefore, the problem should be formulated in a conditional form to make any sense. Then it would be possible to define it as: is it possible to prove that the hypothetical probability distribution P(X1,X2,Y1,Y2) does not exist using only real experimental data of distributions (X1,Y1), (X1,Y2), (X2,Y1) and (X2,Y2)? The conditional form of the formulation is due to the fact that we do not directly imply inconsistency of sets (X1,X2) and (Y1,Y2). Nonetheless, it is possible to answer the question even in this formulation. The contradiction between quantum mechanics and Bell's inequality demonstrates that the hypothetical probability distribution has no physical meaning even as an abstract object.

As the result, the assumption of hidden variables, introduced to combine inconsistent variables, proves invalid given the real data of quantum mechanics.

### 2.5. Problem of existence of hypothetical combined distribution for incompatible observables without use of inequalities. Cochen-Specker model.

To get a classical probability space one needs to define three objects ($\Omega, F, P$), where $\Omega$ - is the sample space, $F$ - is the algebra of events, ($\sigma$ - algebra of subsets of $\Omega$), $P$ - probability of events. Previously it was demonstrated that in Bell's formulation, the proof of impossibility of a hypothetical joint distribution for inconsistent observables leads to a contradiction between measurement results and some inequality describing non-negative definiteness of a probability distribution. In this case inconsistency of a classical probability space is due to impossibility of adequate definitions of probabilities $P$ (some of the probabilities take negative values).

However, it would be interesting to construct models, where the inconsistency of a classical probability space would lead to a logical contradiction during the very first step of sample space $\Omega$ construction. Let us consider two such models. The first one was introduced as early as in 1967 in Cochen-Specker paper [20]. The other model is concerned with highly-relevant GHZ-states and is considered in the next chapter.

Let us consider a particle with spin $s = 1$. We shall use the following simple matrices as spin projection operators [21]:

$(s_j)_{kl} = -i\varepsilon_{jkl}$ ; $\quad j,k,l = 1,2,3$ ,

Whare $\varepsilon_{jkl}$ - is a completely anti-symmetric tensor.

In extended form these matrices are as follows:

$$s_1 = \begin{pmatrix} 0 & 0 & 0 \\ 0 & 0 & -i \\ 0 & i & 0 \end{pmatrix} ; \ s_2 = \begin{pmatrix} 0 & 0 & i \\ 0 & 0 & 0 \\ -i & 0 & 0 \end{pmatrix} ; \ s_3 = \begin{pmatrix} 0 & -i & 0 \\ i & 0 & 0 \\ 0 & 0 & 0 \end{pmatrix}$$

Note that there are also other representations that may we used (e.g. as in the actual Cochen-Specker paper ([20], c. 71)).

The squared momentum operator is:

$s^2 = s_1^2 + s_2^2 + s_3^2 = 2I$,

Where $I$ - is a unitary matrix.

Momentum projections (eigenvalues of operators $s_1$, $s_2$, $s_3$ ) can take three values: -1, 0 и 1 and, therefore, the squares of momentum projections can take either 0 or 1.

Matrices $s_1$, $s_2$, $s_3$ satisfy the following commutation relations:

$$[s_j, s_k] = s_j s_k - s_k s_j = i\varepsilon_{jkl} s_l \ ; \quad j,k,l = 1,2,3$$

The main idea of the Cochen-Specker model is based on the remarkable feature that the squares of operators of momentum projection commute with each other for a unitary spin particle.

$$[s_k^2, s_l^2] = 0 \ ; \quad k,l = 1,2,3$$

It implies that observables $s_1^2$, $s_2^2$, $s_3^2$ can be simultaneously defined. Also, according to equation $s(s+1) = 2$, the sum of the three values is equal to 2, which implies that two of the observables should take the value of 1 and the third observable the value of zero.

The measurement results can be vividly presented by colors on the measurement axes. Let us paint two of the directions corresponding to the value of 1 in red and the third one which corresponds to zero in green. Then we shall take one of the primary basis vectors and two new vectors to form a new basis. As the result, we would get two distinct bases with one mutual vector: $s_1^2$, $s_2^2$, $s_3^2$, and $s_1^2$, $s_2'^2$, $s_3'^2$. Here observable $s_1^2$ is mutual for the two bases, while pairs ($s_2^2$, $s_3^2$) and ($s_2'^2$, $s_3'^2$) are inconsistent (do not commute). Bells problem is concerned with an implicit construction of a joint distribution (classical probability space) for inconsistent (non-commuting) observables. Let us discuss the motivation behind the problem. In this example we can see that two incompatible sets ($s_1^2$, $s_2^2$, $s_3^2$) and ($s_1^2$, $s_2'^2$, $s_3'^2$) become implicitly combined into a joint set by means of a mutual observable $s_1^2$. Then the question whether it is possible to construct a hypothetical joint distribution $P(s_1^2, s_2^2, s_3^2, s_2'^2, s_3'^2)$ for the values naturally arises. Note here that for the sake of convenience, we use same notations to denote observables' operators and measurement results. For example, $s_1^2$ in $P(s_1^2, s_2^2, s_3^2, s_2'^2, s_3'^2)$ stands for a measurement result (either 0 or 1), while in some other formulas it is an operator.

In our color interpretation, combining different sets of observables implies to paint vectors in the manner that the color of direction $s_1^2$ in a new set ($s_1^2$, $s_2'^2$, $s_3'^2$) is the same as in the initial set ($s_1^2$, $s_2^2$, $s_3^2$). At the same time it is clear that we may conduct either measurements ($s_1^2$, $s_2^2$, $s_3^2$) or ($s_1^2$, $s_2'^2$, $s_3'^2$) on any particle. Therefore, if say we conducted measurements ($s_1^2$, $s_2^2$, $s_3^2$), then the results of measurements ($s_1^2$, $s_2'^2$, $s_3'^2$) are treated as some latent (hidden) gist that would have been observed if only we had chosen measurements ($s_1^2$, $s_2'^2$, $s_3'^2$). In Bell's terminology it implies that the Cochen-Specker model implicitly assumes that condition of non-contextuality is met (the color of $s_1^2$ is not dependent on the context, i.e. the chosen set of measurements). Note that in quantum mechanics there are no grounds to believe that such a condition is met. In this sense, the condition of context-independence is merely a kind of metaphysical assumption that is either necessary or unacceptable depending on which philosophy we adopt (see chapter 4 below).

Here we can again note that in accordance with von Neumann's beliefs, the question of convergence of a quantum statistics to classical probability leads to the question of possibility of some imaginary construction. The latter question has already been formulated above in the following form: how can a joint probability distribution for inconsistent observables be constructed? The very formulation of the question, as well as quantum physics in general (theory and experiment), are crying that it can not be done. However, we may formulate the question in another way: can we construct such a distribution as a fiction, which, however, would not embed any internal contradictions? To avoid obvious contradictions between quantum mechanics and experiment in such an imaginary construction we shall agree in advance to not jointly consider non-commuting observables (say $s_2^2$ and $s_2'^2$) directly (in one measurement), but rather combine them indirectly (through measurements of sets ($s_1^2$, $s_2^2$, $s_3^2$) and ($s_1^2$, $s_2'^2$, $s_3'^2$)).

A very interesting result has been obtained using such an indirect formulation by Bell (in 1964 and 1966 papers) and Cochen-Specker (in 1967). It appears that a joint distribution of incompatible observables is impossible even as an imaginary fiction, because it contains intrinsic contradictions. In the color scheme presented above, this implies that we will necessary come to a contradiction after considering a sufficiently large number of bases (when some direction will have to be painted in both red and green). Cochen and Specker constructed a relatively complex graph with 117 nodes. A more simple proof of the result is presented in Mermin paper [22]. In the next chapter we shall discuss the result with regards to a highly important example of Greenberger- Horne- Zeilinger states.

### 2.6. Greenberger- Horne- Zeilinger states.

The results by Bell and Cochen and Specker can be vividly demonstrated using the example of three-particle states (so called Greenberger- Horne- Zeilinger (GHZ) states). [23- 25]. It is convenient to follow Mermin's approach here [22].

There are 10 operators presented on Fig. 7. The operators are considered in a 8D Hilbert space, constructed by a system of three qubits.

For example, $\sigma_x^1 = \sigma_x \otimes I \otimes I$ is operator $\sigma_x$, acting on the first particle, $I$ - is a unitary operator. Similarly, $\sigma_x^2 = I \otimes \sigma_x \otimes I$ is operator $\sigma_x$, acting on the second particle. All six one particle operators $\sigma_x$ and $\sigma_y$ are presented on the figure. There are also four three-particle operators, e.g. operator $\sigma_y^1 \sigma_y^2 \sigma_x^3 = \sigma_y \otimes \sigma_y \otimes \sigma_x$, corresponding to $\sigma_y$ acting on the first and the second particle, and $\sigma_x$ on the third particle.

The ten operators form a star that can be divided into 5 sets of four operators each. Each set forms a line between two of the nodes. It is not difficult to check that all four operators in each sets commute with each other. Each operator lies at the intersection of two edges of the star and therefore belongs to two different sets simultaneously.

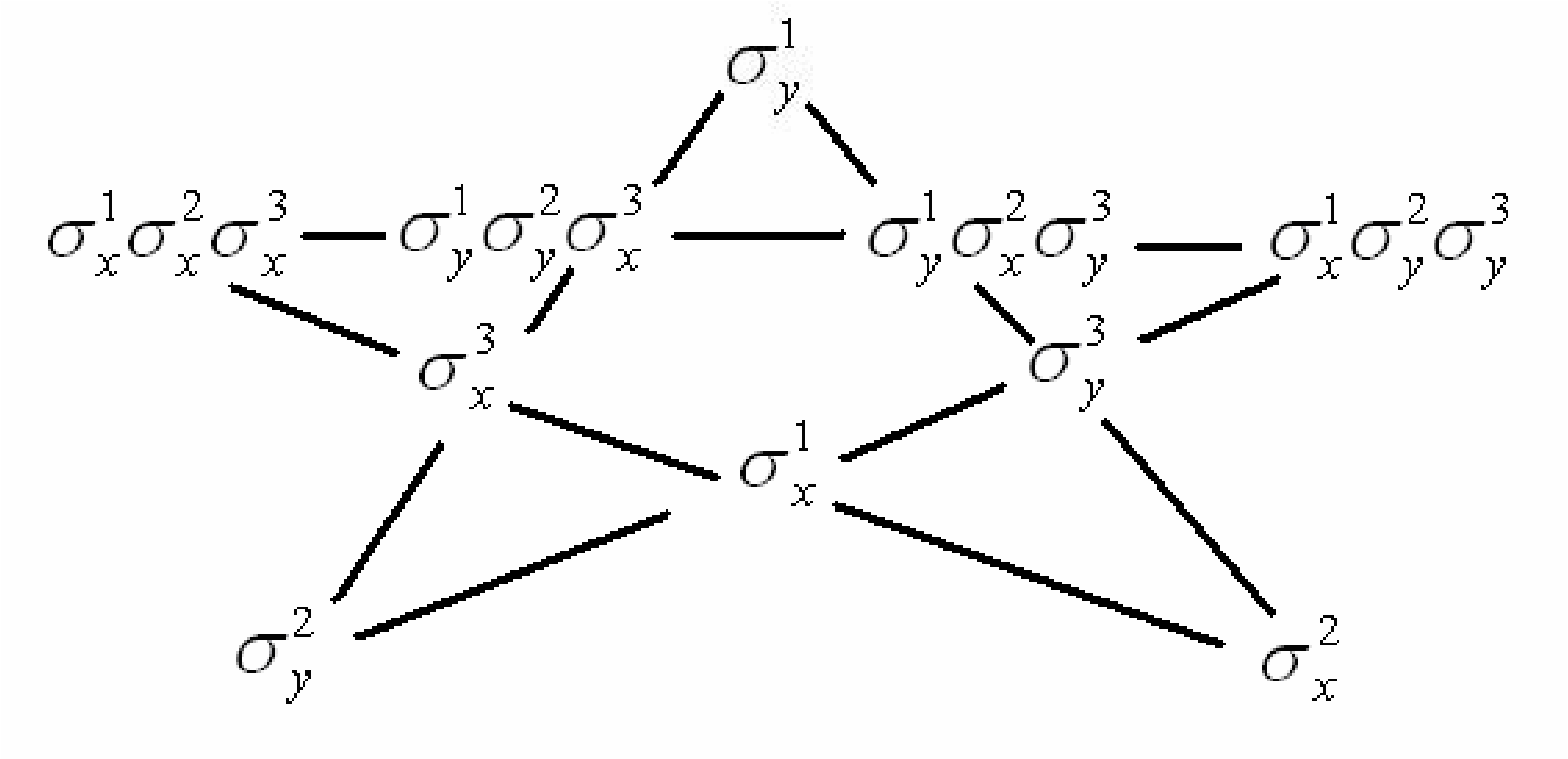

Fig. 7. Three particle measurement scheme.

Each of the observables can take either -1 or +1. It is clear that the total set of 10 operators is incompatible. It can be shown that an attempt to implicitly combine the observables, analogous to painting vector triples in the initial problem of Cochen-Specker, ia also impossible as it leads to a contradiction.

In fact, the product of four measurements on each edge except the horizontal one, is equal to +1, while it is equal to -1 for the horizontal edge. Therefore, the product of all observables should be equal to -1. However, in the product of all the observables, each operator is included twice, and thus the product is equal to +1. The contradiction demonstrates that the observables are incompatible even if combined indirectly.

Let us demosnstrate that the result obtained by the authors has statistical nature behind it. In the framework of statistics, we can see here an attempt to build a classical sample space $\Omega$, consisting of $2^{10} = 1024$ elements. Greenberger, Horne and Zeilinger, in fact, proved that each element of $\Omega$ has zero probability. However, if that is the case then complete probability will be equal to zero rather than one, which leads us to a confusion.

Let us explain how zero probability of a set follows from impossibility of its simultaneous hypothetical realization. Consider an arbitrary set of 10 numbers, where each number is equal either to +1 or -1. The set will contradict with quantum mechanics in at least one of five edges of the star. Such an edge corresponds to a real physical experiment, where real (not hypothetical) probability of joint distribution of the four compatible observables on the edge should be equal to zero. This zero probability consists, in its turn, of 64 hypothetical elementary probabilities (all possible combinations of 6 remaining observables $2^6 = 64$). All of the probabilities should be equal to zero (including the probability corresponding to the initial set of 10 numbers). In other words, each of the 1024 hypothetical probabilities is a summand in the total probability, which is equal to zero, which implies that each summand also has zero value.

For instance, consider a set corresponding to a real experiment ($\sigma_x^1$, $\sigma_y^2$, $\sigma_y^3$, $\sigma_x^1\sigma_y^2\sigma_y^3$). It will be intrinsically contradictory, when one of the observations contradicts with three of the others, e.g. when ($\sigma_x^1 = 1$, $\sigma_y^2 = 1$, $\sigma_y^3 = 1$, $\sigma_x^1\sigma_y^2\sigma_y^3 = -1$). Such a combination has zero probability. In total, there are 16 probabilities in this real experiment with four observables, and 8 of the probabilities are a priori equal to zero. Matrix $A$ in equation $Ap = b$ from chapter 2.2 now consists of 80 rows (5 sets of 16 measurable probabilities) and 1024 columns (the number of hypothetical elementary probabilities). Each row of the matrix contains $2^6 = 64$ elements equal to one and 1024 – 64 = 960 zeros. Column $b$ consists of 80 elements, half of which are a priori equal to zero. Each of the 1024 elementary probabilities, which form column $p$ has to be included in at least one of the rows with zero right-side.

The main idea of the discussion is very simple. There are some cases, when it is possible to derive 64 variables from a single linear equation (when the right side is equal to zero, and the left side is the sum of 64 unknown probabilities.)

GHZ states discover a more profound contradiction between classical and quantum statistics than Bell's inequality does. In Bell's case the contradiction is apparent only for some of the states, while in the case of GHZ it is true for all of the GHZ states.

**2.7. Gleason's theorem, complementarity principle and hidden variables.**

It is a well-known fact in quantum mechanics that von Neumann's projection measurement is defined by a complete set of orthonormal projection operators $\hat{P}_i \quad i = 1,...,s$ that are in accordance with the following conditions:

$$\sum_{i=1}^{s} \hat{P}_i = 1 \qquad \hat{P}_i^2 = \hat{P}_i \qquad \hat{P}_i \hat{P}_j = 0 \text{ при } i \neq j$$

Here $s$ - is the Hilbert space dimension

The projection measurement defines probability distribution (set of non-negative values normalized by unity)

$$p_i \quad i = 1,...,s, \; \sum_{i=1}^{s} p_i = 1$$

Another set of projection operators $\hat{P}'_i \quad i = 1,...,s$ defines some other probability distribution that is complementary to the initial one:

$$p'_i \quad i = 1,...,s$$

Inside of each set of the operators $\hat{P}_i$ and $\hat{P}'_i$ all the operators are consistent with each other (due to the condition $\hat{P}_i \hat{P}_j = 0$ for $i \neq j$). However, the sets of operators $\hat{P}_i$ and $\hat{P}'_i$, considered as a whole are non-consistent with each other (some of the operators $\hat{P}_i \quad i = 1,...,s$ are inherently not consistent with some of the operators $\hat{P}'_i \quad i = 1,...,s$). In a 2D-Hilbert space, the two sets of projection operators $\hat{P}_i$ and $\hat{P}'_i$ do not intersect (due to the transitiveness of the consistency property – see chapter 2.4). This case if of little interest for us, as the different measurement sets are not interconnected. Therefore, we shall suppose that the Hilbert space dimension is not less than three (then the transitivity property does not hold).

Let us consider all possible sets of projection operators $\hat{P}_i$, $\hat{P}'_i$, $\hat{P}''_i$ etc. The set is not only infinite, but also continuous. We shall associate some set of mutually-non-contradictory probability distributions with the projection operators sets. The condition of mutual consistency of the distributions is not so trivial as it may seem, because the different distributions may intersect (same probabilities may belong to different probability distributions).

In mathematics, such a set of mutually- non-contradictory distributions is referred to as a probability measure defined on the set of projectors in Hilbert space [26, 27].

The Gleason's theorem states that in Hilbert space of dimensions three and higher, the set of mutually-non-contradictory distributions unambiguously defines a single density matrix $\hat{\rho}$ so that:

$$p_i = Tr\left(\hat{\rho}\hat{P}_i\right)$$

A more rigorous formalized description of the theorem is given in [26, 27].

The equation $p_i = Tr\left(\hat{\rho}\hat{P}_i\right)$ is a fundamental result of quantum mechanics. It has been derived by von Neumann and generalizes the well-known Born formula. The equation states that for a known density matrix one may calculate the probabilities of any projection measurement. The Gleason's theorem may be thought of as the inverse to the Born and von Neumann's theorems (if we manage to build a combination of mutually non-contradictory distributions on the set of projectors, then the combination would unambiguously define some single theoretical density matrix). We shall interpret the set of mutually non-contradictory distributions as results of quantum measurements. In the interpretation, Gleason's theorem states the possibility of solving the inverse problem of reconstruction of unknown quantum states by results of measurements. In the context of Bohr's complementarity principle, we may call the set of mutually non-contradictory distributions as the set of mutually-complementary distributions.

When considering the theorems of Born-von Neumann and Gleason together, they describe statistical meaning of complementarity principle: a quantum state defines a set of different mutually-complementary probability distributions and the other way around – we may reconstruct quantum state density matrix by known mutually-complementary probability distributions. The term mutually-complementary distributions implies that the distributions complement each other in the sense that taken together, they define some quantum state [16-18].

Complementarity principle describes some isomorphism between quantum theory and experimental studies of quantum events. It appears that a quantum state defined by a state vector or a density matrix, can be experimentally measured by the set of mutually-complementary distributions. The idea of any theory is to provide probabilities for measurements' outcomes. And the other way round, in any correct experiment dealing with a quantum ensemble, results of all possible measurements can be described in a non-contradictory and unambiguous manner.

Complementarity principle is a theoretical basis of quantum tomography, which attempts to reconstruct a quantum state by statistical data, derived in various mutually-complementary experiments. However, note that Gleason's theorem provides the basis for quantum tomography only in principle, because it merely states existence of a density matrix, when all possible mutually-complementary measurements are conducted (in practice, we need defined algorithms to reconstruct a density matrix by a finite set of data).

In a Hilbert space of dimension $s$, a density matrix is defined by a finite number of real parameters equal to $s^2 - 1$. Therefore, to reconstruct the density matrix we shall need the same number (or a bit larger, but still finite number) of mutually-complementary measurements. The set of mutually-complementary measurements does not have to be finite, but then existence of the density matrix should be defied a priori. In experimental studies the a priori existence is guaranteed by quantum mechanics (and of course by experiment's accuracy).

The a priori guarantee of existence of a density matrix is of much importance. That is because if the set of mutually non-contradictory distributions is finite, then the Gleason's theorem on its own can not guarantee existence of a single density matrix.

The Gleason's theorem greatly simplifies the proof of impossibility of leading quantum statistics to a classical probability distribution. Due to condition $\hat{P}_i^2 = \hat{P}_i$, projection operators can take only two eigenvalues: 1 and 0. In a hypothetical classical sample space, we should associate combinations of possible sets ( $\hat{P}_i$, $\hat{P}'_i$, $\hat{P}''_i$ …) with observables $\hat{P}_i$. We shall put in correspondence with observables $\hat{P}_i$ its eigenvalues 0 or 1, and with sets of possible combinations of a single unitary value and zeros. Each point in the hypothetical sample space is a combination of all possible mutually non-contradictory distributions containing one unitary value and many zeros. Then by Gleason's theorem, such a combination of all possible mutually non-contradictory distributions should unambiguously define some density matrix $\hat{\rho}$. Equation $Tr\left(\hat{\rho}\hat{P}_i\right)$ should yield either 0 and 1 for any projector from all possible sets ( $\hat{P}_i$, $\hat{P}'_i$, $\hat{P}''_i$ etc.), which is obviously not possible for any density matrix $\hat{\rho}$. This contradiction demonstrates the impossibility to define a hypothetical classical sample space in a non-contradictory manner. Similar to the Cochen-Specker color problem, we shall inevitable be forced to assign two distinct values (zero and one) to some operator $\hat{P}_0$.

Finally, note the similarity with the problem of an implicit construction of a classical sample space for quantum observables considered in chapter 2.4. It has been stated above that the problem made sense only for Hilbert spaces of dimensions three and higher, and the same restrictions are applicable in the case of Gleason's theorem.

## 3. Statistical completeness of quantum mechanics.

### 3.1. Incompleteness of classical statistical description.

In the framework of classical probability theory, let us consider variables $x_1, x_2, ..., x_s$ that are have a joint probability distribution $P(x_1, x_2, ..., x_s)$. Existence of such distribution does not exclude the possibility of existence of $r$ complementary variables $x_{s+1}, x_{s+2}, ..., x_{s+r}$ that may be statistically dependent on the former variables. The set of variables is statistically dependent if joint distribution of dimension $s+r$ is non-separable, i.e. it can not be factorized as a product of distributions of dimensions $s$ and $r$

$$P(x_1, x_2, ..., x_s, x_{s+1}, ..., x_{s+r}) \neq P(x_1, x_2, ..., x_s) P(x_{s+1}, x_{s+2}, ..., x_{s+r}).$$

Less formally, this property implies that any connections between initial variables $x_1, x_2, ..., x_s$ may appear to be a fiction, because real physical reasons may be defined not by initial but by complementary hidden variables $x_{s+1}, x_{s+2}, ..., x_{s+r}$. Therefore, by large, any classical statistical analysis may not claim to derive objective scientific conclusions.

As far as one hundred years ago Bernard Shaw laughed at such a situation by saying that statisticians could easily prove that wearing cylinder hats extended life and provided immunity against diseases. This intrinsic drawback of classical statistics is well known [31], and proper researchers consider statistical analysis only as a complementary tool for cause-effect analysis of events.

The question directly relates to the subject of hidden variables. If formal statistical analysis (e.g. correlations) lead to absurd results, then we should address subject analysis (physical, technical, medical, social, etc.) that would reveal the variables that were not accounted for before (hidden variables). For instance, let us seriously consider the Shaw's anecdotic example of wearing cylinder hats. It is quite plausible that 100 years ago the people who used to wear cylinder hats did in fact suffer from tuberculosis less than those who used to wear caps. Then statistically we would observe a high correlation between the types of hats and tuberculosis illnesses. The phenomenon is however ridiculous from the medical point of view. Therefore, let us extend the number of variables from two to three adding a new variable "rich-poor" to existing variables "cylinder-cap" and "ill-healthy". Then we will effectively add the third dimension (wealth) to the former 2D-space (hats-health). There are $2^3 = 8$ points in the new space, and each person in our sample could be described by one of the points. Then suppose that data clusters near two points "poor-ill-cap" and "rich-healthy-cylinder". If we now consider the correlation between "health" and "hats", but excluding the effect of variable "wealth" (partial correlation coefficient [31]), then it will appear that the correlation disappears. Thus, the previous correlation was a fiction that arised because of the third (hidden) variable.

Therefore, it appears that search for new variables is relevant and even desirable. However, inclusion of new variables would lead to an increase in sample space dimension, which would complicate statistical analysis until it becomes impossible because of the huge sample size needed. This phenomenon is called "curse of infinity ".

Classical objects (contrary to quantum objects) are treated as informationally unlimited, and therefore no matter how many variables we already included, there will inevitably be an infinite number of unaccounted (hidden) variables. Thus, classical statistics is doomed to remain incomplete.

### 3.2. Completeness of a pure quantum state

It is remarkable that quantum theory does not have the drawbacks analogous to the ones described above. For instance, let variables $x_1, x_2, ..., x_s$ form a pure quantum state $\psi(x_1, x_2, ..., x_s)$. Then, abstractly speaking, the possibility of statistical dependence of the variables on any other variables in the universe is excluded (including hidden variables in the system). In other words, expansion of the initial system by introducing any other additional variables $x_{s+1}, x_{s+2}, ..., x_{s+r}$ will definitely result in a separable joint quantum state, so that the joint quantum state could be represented as a tensor product of independent state vectors $\psi(x_1, x_2, ..., x_s, x_{s+1}, ..., x_{s+r}) = \psi^{(1)}(x_1, x_2, ..., x_s) \otimes \psi^{(2)}(x_{s+1}, x_{s+2}, ..., x_{s+r})$ (e.g. when introducing spin into a non-relativistic quantum mechanics, a state vector becomes a product of coordinate and spin functions). Actually, we statistically described a phenomenon of elementary quantum mechanics. It is well known that to define a quantum state, one needs to describe a complete set of consistent observables $\hat{x}_1, \hat{x}_2, ..., \hat{x}_s$. If we remove at least one operator from the list, then the set loses its completeness. From the other side, the set can not further completed, because no operator commuting with all other already included operators exists. We can define a quantum state, where all observables of the complete set are simultaneously defined, and $\psi(x_1, x_2, ..., x_s)$ is a superposition of all such states. We can define another complete set, e.g. $\hat{y}_1, \hat{y}_2, ..., \hat{y}_s$. A new wave function $\psi(y_1, y_2, ..., y_s)$, however, will be equivalent to the initial one $\psi(x_1, x_2, ..., x_s)$ (because they are interconnected by an a priori known unitary transformation). This property demonstrates that wave function $\psi(x_1, x_2, ..., x_s)$ completely describes a quantum system in the sense that it cannot be further completed by introducing new variables.

Now let us suppose that the state is non-separable, i.e. $\psi(x_1, x_2, ..., x_s, x_{s+1}, ..., x_{s+r}) \neq \psi^{(1)}(x_1, x_2, ..., x_s) \otimes \psi^{(2)}(x_{s+1}, x_{s+2}, ..., x_{s+r})$. Then it is impossible to assign any state vectors to subsystems $x_1, x_2, ..., x_s$ and $x_{s+1}, x_{s+2}, ..., x_{s+r}$. Such subsystems can not be considered as independent closed systems, no matter how far they are from each other. A well known example of such kind of systems is an EPR-state (e.g. a two-qubit singlet state described above). Thus, the notion of closeness in quantum theory is substantially different from analogous notion in classical theory. Spatial isolation can no longer serve as a feature of closeness. Instead, there is a specific statistical criterion in quantum theory: a complete internally closed description, independent from values of any other variables, is possible only for quantum systems that can be described by a state vector. Therefore, EPR states contrary to authors' intention, serve as an important argument in favor, rather than against of completeness of a statistical model of quantum theory. The property demonstrates that a pure quantum state can not separated into pieces with some quantum states assigned to each of the parts (some state vector). Here we again observe a radical difference from the classical case. Therefore, quantum states, contrary to classical states, have some property that we may call completeness [26]. Here, it is especially important for us that completeness of a pure quantum state is a feature of its statistical completeness.

The considerations allow us to talk about incompleteness of classical probability theory and completeness of quantum statistics. Note that incompleteness of axioms of classical probability theory is well known (see [2,5] for example). However, an incomplete description is often in quantum theory as well by a so-called density matrix. The density matrix is important, because a system may interact in a complex manner with its environment. Note, however, that formally any density matrix may be completed to a pure state.

**3.3. Mixed states as tools for incomplete description of subsystem.**

Let a quantum state depend on two groups of variables $x_1$ and $x_2$. Below we shall consider conditions under which variables $x_1$ form a complete set defining some pure quantum state regardless of existence of additional variables $x_2$.

A density matrix corresponding to variables $x_1$ and $x_2$ is as follows:

$$\rho(x_1, x_2; x_1', x_2') = \psi(x_1, x_2)\psi^*(x_1', x_2')$$

Taking an integral on variables $x_2$, we shall get a density matrix of subsystem $x_1$.

$$\rho(x_1; x_1') = \int \psi(x_1, x_2)\psi^*(x_1', x_2')dx_2$$

Let us present a wave function of a system in form of Schmidt decomposition.

$$\psi(x_1, x_2) = \sum_j \sqrt{\lambda_j}\varphi_j^{(1)}(x_1)\varphi_j^{(2)}(x_2),$$

where $\lambda_j$ - are weights, $\sum_j \lambda_j = 1$

Here $\varphi_j^{(1)}(x_1)$ and $\varphi_j^{(2)}(x_2)$ - are so-called Schmidt modes that form orthonormal bases. Then the density matrix of subsystem $x_1$ may be re-written as:

$$\rho(x_1; x_1') = \sum_j \lambda_j \varphi_j^{(1)}(x_1)\varphi_j^{*(1)}(x_1')$$

The result demonstrates that a density matrix is factorisable only if one of the weights is equal to unity and all others to zero ($\lambda_0 = 1$, $\lambda_1 = \lambda_2 = ... = 0$). Then the subsystem state becomes pure and the wave function is represented as a product of wave functions of subsystems $x_1$ и $x_2$:

$$\psi(x_1, x_2) = \varphi_0^{(1)}(x_1)\varphi_0^{(2)}(x_2)$$

As the result, $x_1$ is a complete set and it can not be further completed by introduction of new variables, if and only if the corresponding state of subsystem $x_1$ is pure.

The weights $\lambda_j$ in Schmidt decomposition are eigenvalues of density matrices (they are equal for both subsystems $x_1$ and $x_2$). Orthonormal sets of wave functions $\varphi_j^{(1)}(x_1)$ and $\varphi_j^{(2)}(x_2)$ are eigenvectors of corresponding density matrices. Imagine that we are dealing with subsystem $x_1$ and we know nothing about subsystem $x_2$. Then we may consider some arbitrary imaginary system $x_3$ instead of the real unknown subsystem $x_2$, and also a set of basis vectors $\varphi_j^{(3)}(x_3)$ corresponding to the new system. Then we may consider some imaginary construction instead of the real unknown wave function $\psi(x_1, x_2)$:

$$\psi'(x_1, x_3) = \sum_j \sqrt{\lambda_j}\varphi_j^{(1)}(x_1)\varphi_j^{(3)}(x_3)$$

Note that with regards to the describing the properties of subsystem $x_1$ the imaginary wave function $\psi'(x_1, x_3)$ is no worse than the real one $\psi(x_1, x_2)$. The procedure is called purification of a mixed state. The ambiguity in choosing the imaginary wave function shows that a mixed state described by a density matrix is a statistically incomplete description compared with state vector description.

Schmidt decomposition allows one to introduce such important characteristics the Schmidt number $K$, Schmidt information $I$ and von Neumann entropy $S$ [33]:

$$K = \frac{1}{\sum_k \lambda_k^2}$$

$$I = \log(K)$$

$$S = -\sum_k (\lambda_k \log \lambda_k)$$

Information $I$, based on Schmidt number characterizes degree of statistical correlation (entanglement) of a system with its environment. [33]. Von Neumann's entropy $S$, as a degree of uncertainty of a quantum system described by a density matrix, is a quantum analog of Boltzmann's classical entropy (later an analogous notion appeared in classical theory of Shannon's information). In the case of a pure state $K = 1$, $I = 0$, $S = 0$. It implies that a quantum system in a pure state is not entangled with its environment, and the information of it can not be completed, because ther is no uncertainty in the state. For a mixed state $K > 1$, $I > 0$, $S > 0$ instead. A quantum system in a mixed state is statistically dependent on its environment and is characterized by uncertainty, which could be eliminated only if we move from a density matrix to a more complete description of the whole system by a state vector.

**4. Psychology and philosophy behind the battle for hidden variables.**

**4.1. Bell's psychological view of von Neumann's model.**

Unfortunately, principal questions of quantum mechanics are not always the subject of unprejudiced scientific analysis. Often, quantum mechanics becomes the issue of intense political debates. In the problem of hidden variables, there has been a long struggle between two parties in favor and against their existence (following Einstein and Bohr correspondingly [3]). However while the historical debate between Einstein and Bohr can serve as an ethical example, it has often not been the case for their followers. For example this is how Bell characterizes von Neumann's proof of impossibility of introducing hidden variables into quantum mechanics:

«Yet the von Neumann proof, if you actually come to grips with it, falls apart in your hands! There is *nothing* to it. It's not just flawed, it's *silly!* ... When you translate [his assumptions] into terms of physical disposition, they're nonsense. You may quote me on that: The proof of von Neumann is not merely false but *foolish!* (Interview in Omni, May, 1988, p.88, cited by the paper of Mermin [22], italics by original text, see also [34]).

Note that Mermin's work contains much valuable material of a summary of models with hidden variables. However, Mermin's complete consent with the rude comments by Bell and others with reference to von Neumann and other founders of quantum mechanics, as well as methodology of quantum mechanics, is regrettable.

Let us however move from the ethical aspect of the discussion to a psychological aspect. The intonation of Bell's comments is such of a person that is absolutely confident in his rightfulness and opponent's falseness.

However, specifics of quantum mechanics is such that under close investigation even the most obvious conclusions may prove to be wrong.

**4.2. So what else is there that Bell did not notice?**

It has been already mentioned in chapter 2.4 that historically the problem of hidden variables emerged with regards to the problem of describing quantum statistics by a classical probability distribution. It has already been noted that the problem has two formulations: direct and indirect one. In the direct approach experimental data and mathematical formalism immediately reject the possibility of classical description of quantum statistics. Bell's results (and his inequalities) refer only to the indirect approach, when we agree not to consider direct incompatibility of non-commuting variables and attempt to combine them in an explicit manner. Analysis of Bell's works demonstrates that he ignores direct formulation of the problem, as well as concrete physical arguments by von Neumann, considering existence of principally incompatible observables. Similar considerations by Bohr, Heisenberg and others are also ignored, which is at the heart of Bell's reluctance to accept such principle notions of quantum mechanics as complementarity principle and uncertainty relations.

Let us again address the assumptions of von Neumann's theorem that are so feverishly attacked by Bell. The analysis presented above demonstrates that the conditions criticized by Bell, which consider analysis of two incompatible observables are perfectly logical for a direct formulation of the problem. Therefore, the whole conflict is due to Bell's reluctance to accept nuances that are inevitably present for different formulations of the same problem, which result in his attempt to present his point of view as the only right one, ignoring arguments of others.

Note that in a straightforward approach even Bell's results appear either as banal or senseless. In chapter 1.4 we already demonstrated that the model with hidden variables for a ½-spin is nothing but a well-known random number generator. Remarks by Wigner, von Neumann's friend prove that von Neumann was well accustomed with the model that Bell discovered more than 30 years after his works. Moreover, von Neumann often used the model, for instance in unpublished discussions with Schrödinger (see [27], appendix).

**4.3. Physics and metaphysics.**

By a bright remark of von Neumann "in the atom we are at the boundary of the physical world" ([6], p. 305). This may be the reason for the close ties between physics and metaphysics (philosophy). However, the mix of such questions into a single conglomerate, when the frontier between natural and humanitarian sciences becomes blurred may be rather harmful for both sciences. For example, let us consider the question of violations of Bell's inequalities in quantum mechanics. A standard formulation (see [35] for example), says that violations of Bell's inequalities implies absence in the Nature of either Einstein's realism or locality, or of both. The main drawback of the formulation is in the unacceptable mixing of precise sciences and philosophy. It seems that we may obtain impressive philosophical conclusions by performing some simple algebraic computations. It appears that physicists are able to transform philosophy into elementary school algebra: multiply different variables, sum some of the results and subtract some of the other and you will be able to distinguish reality and unreality. At the same time poor philosophers do not know of the discovery and continue to doubt everything following Kant: “What can I know? What should I do? What can I hope for?”

Conscious and insisting aspiration to combining physics and metaphysics is in fact quite common for Bell and its followers. M. Scully attributes the following citation to Bell: It would be interesting if studies of quantum mechanics could lead us to the proof of existence of the God or Buddha ([36]). It is worth thinking whether such approach sends physics to the level of medieval scholasticism. It is interesting to know the value of such propositions from the point of view of theology. Does the position remind the Biblical character Thomas? Jesus said to Thomas, "Is it because you have seen me that you have believed? How blessed are those who have never seen me and yet have believed!" **(**John 20:29**)**.

In chapters 2.1 and 2.2 we have already seen that violations of Bell’s inequalities in quantum mechanics has a natural scientific and not philosophical nature. No imaginary (hidden) probability distribution P(X1,X2,Y1,Y2) for a set of incompatible observables exists. Existence of such distribution we assumed during transition from Bell-Clauser-Horne-Shimony-Holt inequality to Bell’s inequality. Thus the paradox of violations of Bell’s inequalities is imaginary. There is no paradox – the erroneous chain if conclusions, which considers Bell’s inequality, can mistreat only a person who does not understand a thing in statistics. Rephrasing a well-known joke we may say: “Bell’s paradox is a tax on those who don’t know statistics” (Originally:  Lottery is a tax on those who don’t know statistics”)

## 5. Discussion and conclusion

### 5.1. Discussion

The analysis presented in this paper demonstrates three main issues arising when considering models with hidden variables.

Determinism vs. randomness.

The problem leads to the following question: is it possible (at least hypothetically) to introduce some hidden variables that would transform random outcomes into deterministic ones in quantum mechanics? By its very nature, quantum mechanics can not give exact predictions for single representatives of an ensemble. This feature allows for construction of a formal model with hidden variables that generalizes quantum mechanics, so that the latter becomes an asymptotical limit of a theory with hidden variables.

Quantum statistics vs. classical probability theory.

The problem refers to the following question: is it possible to describe quantum effects using classical probability distributions? The question may be addressed either directly or indirectly. For the indirect approach, there is some hypothetical distribution that indirectly combines incompatible observables. Still, both approaches lead to impossibility of describing quantum statistics by means of classical statistics.

Completeness vs. incompleteness of quantum mechanics.

The analysis allows one to refer to dynamic Gödel -type incompleteness and statistical completeness of quantum mechanics. Dynamic incompleteness is inherent to any statistical description (both classical and quantum), while statistical completeness is a special feature of quantum mechanics and it is not observed in classical theory. An example of statistical completeness of quantum mechanics is the fact that a full set of compatible observables can not be completed neither directly nor indirectly. A pure quantum state has internal completeness and closeness that a mixed state of a quantum system entangled with its environment lacks.

### 5.2 Conclusion

Let us briefly summarize the main contributions of the work:

1. A quantum model based on hidden variables, which generalizes the Bell-type model for a single spin, is proposed and analyzed. Dynamics of hidden variables during measurements leading to dynamic chaos is considered. It is demonstrated that the equilibrium state that the system asymptotically evolves is in perfect agreement with results of quantum theory.

2. Using svd-decomposition, analysis of violations of Bell's inequalities in view of existence of negative probabilities is performed. The analysis which completes works by Sudarshan and Rothman allows to present all solutions in a compact format. A regularized solution, which minimizes dispersion is obtained; an invariant which highlights the fact of violations of Bell's inequalities is proposed.

3. The thesis of incompleteness of quantum mechanics in a dynamic interpretation and its completeness in a statistical interpretation is substantiated. Arguments in favor of incompleteness of a classical statistical description are provided.

4. Conceptual and philosophical questions regarding completeness of quantum mechanics, comparisons between classical and quantum statistical models and existence of hidden variables have been discussed.